\documentclass[twocolumn]{aastex631}

\usepackage{amsmath}
\usepackage{color}
\usepackage{graphicx}
\usepackage{graphbox}
\usepackage{natbib}
\usepackage{booktabs}
\usepackage{xspace}
\usepackage{array,multirow}
\usepackage{amsmath}
\usepackage{wrapfig}

\newcommand\mc[1]{\multicolumn{1}{c}{#1}}

\newcommand{\oj}{OJ\,287\xspace}
\newcommand{\ra}{\emph{RadioAstron}\xspace}

\mathcode`*=\string"8000
\begingroup
\catcode`*=\active
\xdef*{\noexpand\textup{\string*}}
\endgroup

\received{7 September 2021}
\submitjournal{ApJ}

\shorttitle{Space and millimeter VLBI imaging of \oj}
\shortauthors{G\'omez et al.}

\graphicspath{{./}{figures/}}

\begin{document}

\title{Probing the innermost regions of AGN jets and their magnetic fields with RadioAstron. \\ 
V. Space and ground millimeter-VLBI imaging of OJ\,287}

\author[0000-0003-4190-7613]{Jos\'e L. G\'omez}
\affiliation{Instituto de Astrof\'{\i}sica de Andaluc\'{\i}a-CSIC, Glorieta de la Astronom\'{\i}a s/n, 18008 Granada, Spain}

\author{Efthalia Traianou}
\affiliation{Instituto de Astrof\'{\i}sica de Andaluc\'{\i}a-CSIC, Glorieta de la Astronom\'{\i}a s/n, 18008 Granada, Spain}
\affiliation{Max-Planck-Institut f\"ur Radioastronomie, Auf dem H\"ugel 69, 53121 Bonn, Germany}

\author{Thomas P. Krichbaum}
\affiliation{Max-Planck-Institut f\"ur Radioastronomie, Auf dem H\"ugel 69, 53121 Bonn, Germany}

\author{Andrei P. Lobanov}
\affiliation{Max-Planck-Institut f\"ur Radioastronomie, Auf dem H\"ugel 69, 53121 Bonn, Germany}
\affiliation{Moscow Institute of Physics and Technology,
  Institutsky per. 9, Dolgoprudny, Moscow region, 141700, Russia}

\author{Antonio Fuentes}
\affiliation{Instituto de Astrof\'{\i}sica de Andaluc\'{\i}a-CSIC, Glorieta de la Astronom\'{\i}a s/n, 18008 Granada, Spain}

\author{Rocco Lico}
\affiliation{Instituto de Astrof\'{\i}sica de Andaluc\'{\i}a-CSIC, Glorieta de la Astronom\'{\i}a s/n, 18008 Granada, Spain}
\affiliation{Max-Planck-Institut f\"ur Radioastronomie, Auf dem H\"ugel 69, 53121 Bonn, Germany}
\affiliation{INAF -- Istituto di Radioastronomia, via Gobetti 101, 40129 Bologna, Italy}

\author{Guang-Yao Zhao}
\affiliation{Instituto de Astrof\'{\i}sica de Andaluc\'{\i}a-CSIC, Glorieta de la Astronom\'{\i}a s/n, 18008 Granada, Spain}

\author[0000-0002-5182-6289]{Gabriele Bruni}
\affiliation{INAF -- Istituto di Astrofisica e Planetologia Spaziali, via Fosso del Cavaliere 100, 00133 Roma, Italy}

\author[0000-0001-9303-3263]{Yuri Y. Kovalev}
\affiliation{Lebedev Physical Institute of the Russian Academy of Sciences, Leninsky prospekt 53, 119991 Moscow, Russia}
\affiliation{Moscow Institute of Physics and Technology,
  Institutsky per. 9, Dolgoprudny, Moscow region, 141700, Russia}
\affiliation{Max-Planck-Institut f\"ur Radioastronomie, Auf dem H\"ugel 69,
53121 Bonn, Germany}

\author{Anne L\"{a}hteenm\"{a}ki}
\affiliation{Aalto University Mets\"{a}hovi Radio Observatory, Mets\"{a}hovintie 114, FI-02540 Kylm\"{a}l\"{a}, Finland}
\affiliation{Aalto University Department of Electronic and Nanoengineering, PL15500, FI-00076 Aalto, Finland}

\author{Petr A. Voitsik}
\affiliation{Lebedev Physical Institute of the Russian Academy of Sciences, Leninsky prospekt 53, 119991 Moscow, Russia}

\author[0000-0001-6088-3819]{Mikhail M. Lisakov}
\affiliation{Max-Planck-Institut f\"ur Radioastronomie, Auf dem H\"ugel 69, 53121 Bonn, Germany}
\affiliation{Lebedev Physical Institute of the Russian Academy of Sciences, Leninsky prospekt 53, 119991 Moscow, Russia}

\author{Emmanouil Angelakis}
\affiliation{Section of Astrophysics, Astronomy \& Mechanics, Department of Physics, National and Kapodistrian University of Athens, Panepistimiopolis Zografos 15784, Greece}

\author{Uwe Bach}
\affiliation{Max-Planck-Institut f\"ur Radioastronomie, Auf dem H\"ugel 69, 53121 Bonn, Germany}

\author{Carolina Casadio}
\affiliation{Institute of Astrophysics, Foundation for Research and Technology, Hellas, Voutes, 70013 Heraklion, Greece}
\affiliation{Max-Planck-Institut f\"ur Radioastronomie, Auf dem H\"ugel 69, 53121 Bonn, Germany}

\author{Ilje Cho}
\affiliation{Instituto de Astrof\'{\i}sica de Andaluc\'{\i}a-CSIC, Glorieta de la Astronom\'{\i}a s/n, 18008 Granada, Spain}

\author{Lankeswar Dey}
\affiliation{Department of Astronomy and Astrophysics, Tata Institute of Fundamental Research, Mumbai 400005, India}

\author{Achamveedu Gopakumar}
\affiliation{Department of Astronomy and Astrophysics, Tata Institute of Fundamental Research, Mumbai 400005, India}

\author[0000-0002-0694-2459]{Leonid I. Gurvits}
\affiliation{Joint Institute for VLBI ERIC (JIVE), Oude Hoogeveensedijk 4, 7991 PD Dwingeloo, The Netherlands}
\affiliation{ Dept. of Astrodynamics and Space Missions, TU Delft, Kluyverweg 1, 2629 HS Delft, The Netherlands }

\author{Svetlana Jorstad}
\affiliation{Institute for Astrophysical Research, Boston University, 725 Commonwealth Avenue, Boston, MA 02215, USA}
\affiliation{Astronomical Institute, St. Petersburg State University, Universitetskij, Pr. 28, Petrodvorets, St. Petersburg 198504, Russia}

\author[0000-0002-8017-5665]{Yuri A. Kovalev}
\affiliation{Lebedev Physical Institute of the Russian Academy of Sciences, Leninsky prospekt 53, 119991 Moscow, Russia}

\author{Matthew L. Lister}
\affiliation{Department of Physics and Astronomy, Purdue University, 525 Northwestern Avenue, West Lafayette, IN 47907, USA}

\author{Alan P. Marscher}
\affiliation{Institute for Astrophysical Research, Boston University, 725 Commonwealth Avenue, Boston, MA 02215, USA}

\author{Ioannis Myserlis}
\affiliation{Instituto de Radioastronom\'{\i}a Milim\'etrica, Avenida Divina Pastora, 7, Nucleo Central, E-18012 Granada, Spain}

\author{Alexander B. Pushkarev}
\affiliation{Crimean Astrophysical Observatory, 98409 Nauchny, Crimea, Russia}
\affiliation{Lebedev Physical Institute of the Russian Academy of Sciences, Leninsky prospekt 53, 119991 Moscow, Russia}


\author{Eduardo Ros}
\affiliation{Max-Planck-Institut f\"ur Radioastronomie, Auf dem H\"ugel 69, 53121 Bonn, Germany}

\author{Tuomas Savolainen}
\affiliation{Aalto University Department of Electronics and Nanoengineering, PL 15500, FI-00076 Aalto, Finland}
\affiliation{Aalto University Mets\"{a}hovi Radio Observatory, Mets\"{a}hovintie 114, FI-02540 Kylm\"{a}l\"{a}, Finland}
\affiliation{Max-Planck-Institut f\"ur Radioastronomie, Auf dem H\"ugel 69, 53121 Bonn, Germany}

\author{Merja Tornikoski}
\affiliation{Aalto University Mets\"{a}hovi Radio Observatory, Mets\"{a}hovintie 114, FI-02540 Kylm\"{a}l\"{a}, Finland}

\author{Mauri J. Valtonen}
\affiliation{Finnish Centre for Astronomy with ESO, University of Turku, FI-20500 Turku, Finland}
\affiliation{Department of Physics and Astronomy, University of Turku, FI-20500 Turku, Finland}

\author{Anton Zensus}
\affiliation{Max-Planck-Institut f\"ur Radioastronomie, Auf dem H\"ugel 69, 53121 Bonn, Germany}

\begin{abstract}
\noindent
We present the first polarimetric space VLBI observations of OJ\,287, observed with \textit{RadioAstron} at 22\,GHz during a perigee session on 2014 April 4 and five near-in-time snapshots, together with contemporaneous ground VLBI observations at 15, 43, and 86\,GHz. Ground-space fringes were obtained up to a projected baseline of 3.9 Earth diameters during the perigee session, and at a record 15.1 Earth diameters during the snapshot sessions, allowing us to image the innermost jet at an angular resolution of $\sim50\mu$as, the highest ever achieved at 22\,GHz for OJ\,287. Comparison with ground-based VLBI observations reveals a progressive jet bending with increasing angular resolution that agrees with predictions from a supermassive binary black hole model, although other models cannot be ruled out. Spectral analyses suggest that the VLBI core is dominated by the internal energy of the emitting particles during the onset of a multi-wavelength flare, while the parsec-scale jet is consistent with being in equipartition between the particles and magnetic field. Estimated minimum brightness temperatures from the visibility amplitudes show a continued rising trend with projected baseline length up to $10^{13}$\,K, reconciled with the inverse Compton limit through Doppler boosting for a jet closely oriented to the line of sight. The observed electric vector position angle suggests that the innermost jet has a predominantly toroidal magnetic field, which together with marginal evidence of a gradient in rotation measure across the jet width indicate that the VLBI core is threaded by a helical magnetic field, in agreement with jet formation models.

\end{abstract}



\section{Introduction}
\label{sec:intro}

The BL Lacerta object \oj \citep[0851+202, $z$=0.306; ][]{1989A&AS...80..103S} is considered as one of the most remarkable examples of an active galactic nucleus (AGN). The 12-year quasi-periodic outbursts in the optical regime, often accompanied by multiwavelength flaring activity, are attributed to the existence of a supermassive binary black hole (SMBBH) system, hidden in its compact center \citep[][and references thereafter]{1988ApJ...325..628S,2016ApJ...819L..37V}. The well-aligned with our line of sight radio jet of \oj has shown very remarkable behaviour in the past. During the period 2004-2006, \cite{2012ApJ...747...63A} reported an erratic jet-swing by almost 100$^{\circ}$, attributed to asymmetric accretion flow to the central engine. Systematic jet-axis rotation has been observed by \citet{2017Galax...5...12C} and \cite{2018MNRAS.478.3199B}, describing a 24$-$30 years rotation cycle of a helical jet, explained by either the Lense-Thirring effect introduced by the wobbling of the accretion disk around the primary black hole (BH), or precession induced by the binary companion \citep{1918PhyZ...19...33T,1918PhyZ...19..156L}. 

The elucidation of the origin and the phenomenology of relativistic jets emanating from SMBBH belongs to the frontiers of modern astronomical research. Some theoretical models suggest that relativistic AGN jets can be driven by strong magnetic fields anchored on the black hole’s accretion disk \citep{1982MNRAS.199..883B}. Other scenarios suggest that jets are driven by the conversion of the rotational energy of the BH to Poynting flux via the open magnetic field lines, which are attached to the BH ergosphere \citep{10.1093/mnras/179.3.433}. In reality, superposition of both mechanisms is also possible \citep[e.g.,][]{2000A&A...358..104C}. Close to the central engine, the plasma flow is accelerated and collimated under the presence of a coiled magnetic field. Under such extreme conditions, the dynamics and the stability of the relativistic plasma flow can be strongly influenced by disruptions of the accretion flow, pressure mismatches between the jet and the ambient medium, as well as within the jet itself, leading to the formation of moving shocks, standing shocks, and instabilities \citep{Gomez_1997,1997MNRAS.288..833K,1998ApJ...493..291B,2003ApJ...585L.109A}.  

Very-long-baseline interferometry (VLBI) is a powerful technique that allows us to achieve the angular resolution needed to resolve and study the fine structure of extragalactic jets. By increasing the observed frequency or the baseline length of the array elements, one can probe jet features located in a region of tens of Schwarzschild radii away from the super-massive black hole (SMBH), and ultimately using millimetre wavelengths \citep{2017A&ARv..25....4B} achieve the necessary resolution that has enabled the Event Horizon Telescope Collaboration to obtain the first image of a black hole shadow in M87 \citep{2019ApJ...875L...1E,2019ApJ...875L...2E,2019ApJ...875L...3E,2019ApJ...875L...4E,2019ApJ...875L...5E,2019ApJ...875L...6E,2021ApJ...910L..12E,2021ApJ...910L..13E}. 
Besides VLBI observations in total intensity, polarimetric observations are essential \citep{2001ApJ...561L.161G}. They constitute a powerful tool for deriving fundamental constraints on jet physics and magnetic field configuration.

The images we obtain via VLBI observations allow us to illustrate the relativistic plasma flow and trace distinct features known as blobs or knots. These features usually emerge from the bright and typically unresolved end of the radio jet, known as the core. The nature of this feature differs from that of the real jet base and it is not always easily determined. As also described in \cite{2008ASPC..386..437M}, what we see as the VLBI core can correspond to the surface where the opacity ($\tau_v$) reaches unity \citep{1979ApJ...232...34B}, or to a standing shock, which is located downstream of the $\tau_v=1$ surface \citep{1988ApJ...334..539D,1995ApJ...449L..19G,2015ApJ...809...38M}. The true nature of \oj VLBI core is still unexplored.

The \ra space VLBI mission \citep{Kardashev:2013wu}, led by the Russian Astro Space Center and the Lavochkin Scientific and Production Association, operated between 2011 and 2019. It featured a 10\,m radio telescope on board of the \textit{Spektr-R} satellite and was equipped with receivers operating 0.32\,GHz (P-band), 1.6\,GHz (L-band, dual polarization), 4.8\,GHz (C-band) and 22\,GHz (K-band, dual polarization). With an apogee of $\sim$350 000\,km, space VLBI observations with \ra are capable of imaging blazar jets in total and linearly polarized intensity with an unprecedented resolution of the order of few tens of microarcseconds ($\mu$as) when observing at the shorter wavelengths \citep[e.g.,][]{G_mez_2016}.

Three Key Science Programmes (KSPs) on AGN imaging have collected data since 2013 to study the launching, collimation, and magnetic field properties of AGN jets (e.g., Zensus AdSPR on strong sources program), while the AGN survey studied the brightness temperature of their cores \citep[e.g.,][]{2016ApJ...820L...9K,2020AdSpR..65..705K}. The \ra Polarization KSP has collected data throughout the whole duration of the space VLBI mission, and is aimed to probe the innermost jet regions and their magnetic field in a sample of the most energetic blazars. Results from the \ra Polarization KSP are reported for 0642+449 in \cite{2015A&A...583A.100L}, BL\,Lac in \cite{G_mez_2016}, 0716+714 in \cite{Kravchenko_2020}, 3C\,345 in \cite{2021A&A...648A..82P}, and 3C\,273 in \cite{2017A&A...604A.111B} and \cite{bruni2021}. Here we present the first \ra observations of \oj performed in 2014 at 22\,GHz, in combination with quasi-simultaneous ground VLBI observations at 15, 22, 43, and 86\,GHz.

The structure of this article is as follows. In Sec.~\ref{sec:obs}, we present the multi-frequency data set and the data reduction techniques; in Sec.~\ref{sec:results} we report on the results from the VLBI study and discuss their implications in Sec.~\ref{sec:discussion}.
Throughout this paper we have adopted the following cosmological parameters: $\Omega_\mathrm{M}=0.27$, $\Omega_{\Lambda}=0.73$, $H_{0}=71$~km~s$^{-1}$~Mpc$^{-1}$ \citep{2009ApJS..180..330K}, which result for \oj in a luminosity distance of $D_{L}=1.577$~Gpc and angular scale of 4.48~pc/mas.

\section{Observations and data analysis}
\label{sec:obs}
\oj was observed with \ra in 2014 as part of our Polarization KSP. In this section we describe the \ra observations and data analysis, as well as other close in time ground-based VLBI observations at multiple wavelengths. These include global millimeter VLBI array (GMVA) observations, archival 15\,GHz VLBA data, and publicly available 43\,GHz data from the VLBA-BU-BLAZAR monitoring program. A summary of the \ra and GMVA observations, as well as the used archival data is provided in Tab.~\ref{table:obs}.

\subsection{\textit{\ra} space VLBI observations at 22~GHz}
\label{Sec:RA_data}

The bright blazar \oj was observed with \ra on 4-5 April 2014 (from 12:00 to 03:45 UT, observing code GA030E) at a frequency of 22.236~GHz combining a 15:45 hours session during the perigee of the spacecraft (total \ra on-source time was 6.3~hours). These imaging  observations were supplemented by 5 short (between half an hour and 2 hours long) sessions on 9, 10, 16, 27 March and 18 April, 2014, within the \ra AGN survey program \citep{2020AdSpR..65..705K}.


The \ra perigee imaging session of OJ~287 in 2014 April 4-5 was carried out with a ground array of 12 antennas, as listed in Tab.~\ref{table:obs}. The long-baseline snapshot sessions were conducted in 2014 March 9 (01:00 to 02:00 UT, at a projected baseline of 15.1~Earth diameters, $D_\earth$), March 10 (01:00 to 02:00 UT, 5.8~$D_\earth$), March 16 (2:15 to 03:00 UT, 15.1~$D_\earth$), March 27 (00:00 to 00:25 UT, 4.6~$D_\earth$), and April 18 (00:00 to 01:00 UT, 9.9~$D_\earth$), all during different orbits of the space radio telescope (SRT) than that of the main perigee imaging session. Fringes to the SRT were found only with the Green Bank telescope for all the snapshot sessions except the one on 2014 March 27, for which fringes were found only to Effelsberg. Signal-to-noise ratio (SNR) values of these long-baseline fringes lie in a range from 10 to 20. This is expected, as baseline sensitivity depends on size of the antennas and baseline projection, resulting in no fringe detection for the smaller antennas in our array in the long-baseline snapshots.

The imaging data were recorded in both, left (LCP) and right (RCP) circular polarizations, with a total bandwidth of 32~MHz per polarization, split into two intermediate frequency (IF) bands. The SRT data were recorded by the \ra satellite tracking station in Pushchino, including extended gaps of approximately one-hour duration to allow for cooling of the on board data downlink radio instrumentation of the {\it Spektr-R} satellite. 
The data of the five snapshot survey sessions were recorded in LCP only.

The imaging data were processed using the \emph{RadioAstron}-dedicated version of the {\tt DiFX} software correlator \citep{2016Galax...4...55B}, developed at the Max-Planck-Institut f\"ur Radiostronomie. After setting the ground stations clocks, we performed fringe-fitting between the largest antennas of the ground array and \emph{RadioAstron}, separately for each scan involving it. Such a process allows to minimize the effects of the spacecraft acceleration terms along the considered orbit segment, by re-centering the signal in the correlation window every few minutes of observations. For scans giving no fringes (on the largest baselines), we estimated clock values by extrapolating them from successful scans on shorter space-baselines. This provides a first order value for fringes delay and rate, that can be later refined with the post-processing data reduction software.

The data of the long-baseline snapshot survey sessions were processed by the software correlator developed at the Astro Space Center of Lebedev Physical Institute in Moscow \citep{2017JAI.....650004L}.

\begin{deluxetable*}{lccc}[t]
\tablecaption{\ra and complementary VLBI observations of \oj in 2014.\label{table:obs}}
\tablewidth{0pt}
\tablehead{
\colhead{$\nu_\mathrm{obs}$} & \colhead{Instrument} & \colhead{Observation date} &  \colhead{Participating antennas} \\
\colhead{[GHz]} & \colhead{} & \colhead{} & \colhead{} \\
\colhead{(1)} & \colhead{(2)} & \colhead{(3)} & \colhead{(4)}
}
\startdata
15 & VLBA        &  5 May & $ \mbox{VLBA}^\mathrm{c}$ \\
22$^\mathrm{a}$ & RadioAstron &  4 April & SRT+KVN+KL+ON+SH+TR+HH+NT+EF+YS+GB \\
43$^\mathrm{b}$ & VLBA        & 3 May & $ \mbox{VLBA}^\mathrm{c}$ \\
86          & GMVA            & 24 May & $\mbox{VLBA}^\mathrm{d}$+PV+EB+ON+PB \\
\enddata
\tablecomments{$^\mathrm{a}$ \ra perigee imaging session; see text for description of complementary \ra long-baseline visibility tracking sessions. $^\mathrm{b}$ Data from the VLBA-BU-BLAZAR monitoring program. $^\mathrm{c}$ North Liberty (NL) did not participate. $^\mathrm{d}$ Excluding Hancock (HN) and Saint Croix (SC) antennas which do not have 86\,GHz receivers. Antennas and VLBI array acronyms: VLBA -- Very Long Baseline Array; KVN -- Korean VLBI Network, comprised of the array elements KT, KU, and KY; KL -- Kalyazin; ON -- Onsala; SH -- Sheshan; TR -- Torun; HH -- Hartebeesthoek; NT -- Noto; EF/EB -- Effelsberg; YS -- Yebes; GB -- Green Bank; PV -- Pico Veleta; PB -- Plateau de Bure. Columns from left to right: (1) Observing frequency, (2) VLBI array, (3) Dates of data collection, (4) Array elements that participated in the observations.}
\end{deluxetable*}

\subsubsection{Initial data processing and imaging}
\label{Sec:Fringe_fitting}

The initial data reduction of the correlated data was performed with NRAO's {\tt AIPS} software package \citep{1990apaa.conf..125G}. This includes \textit{a-priori} amplitude calibration using the measured system temperatures for the ground antennas and the SRT. Opacity corrections as a function of source elevation were introduced for the ground telescopes, as well as parallactic angle corrections. 

Fringe fitting of the imaging data was performed first by solving for the instrumental phase and single band delays of the SRT on a scan near the perigee of the SRT, providing the best fringe solutions for the ground-space baselines. This was followed by a global fringe search to solve for the residual delays and rates of the ground antennas only. Once the ground array was calibrated, fringe fitting of the SRT was performed by performing a baseline stacking of the ground array, and setting an exhaustive baseline search with the most sensitive ground antennas. Both polarizations and IFs were combined to maximize the sensitivity for ground-space fringe detection.

\begin{figure}
\centering
\includegraphics[width=0.45\textwidth]{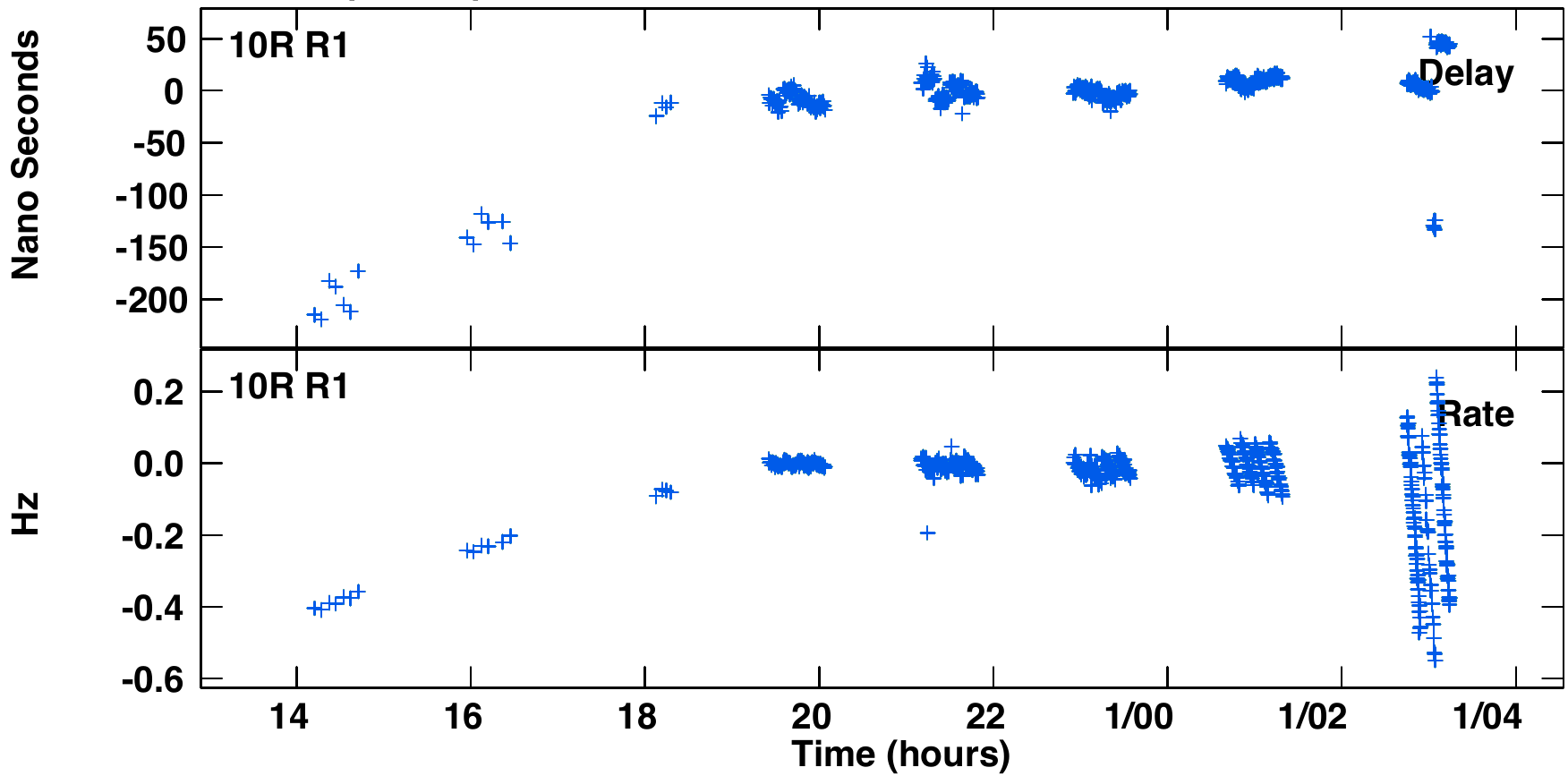}
\caption{Residual delay (top) and rate (bottom) solutions for the SRT during the perigee imaging session on 2014 April 4-5. Note the rapidly changing rate solutions associated with the acceleration of the SRT during the perigee near the end of the observations.}
\label{Fig:FFT_res}
\end{figure}

Figure \ref{Fig:FFT_res} shows the residual delay and rate fringe solutions for the SRT corresponding to the perigee imaging session. The strong ground-space fringes present during the perigee passing of the SRT near the end of the observations allowed for solution intervals as short as 10 seconds, capturing the quickly evolving fringe rates due to the SRT acceleration. Progressively larger solution intervals, up to four minutes, were used to increase the sensitivity on the longer projected baseline lengths to the SRT at the beginning of the experiment.

\begin{figure}
\centering
\includegraphics[align=c,width=0.49\textwidth]{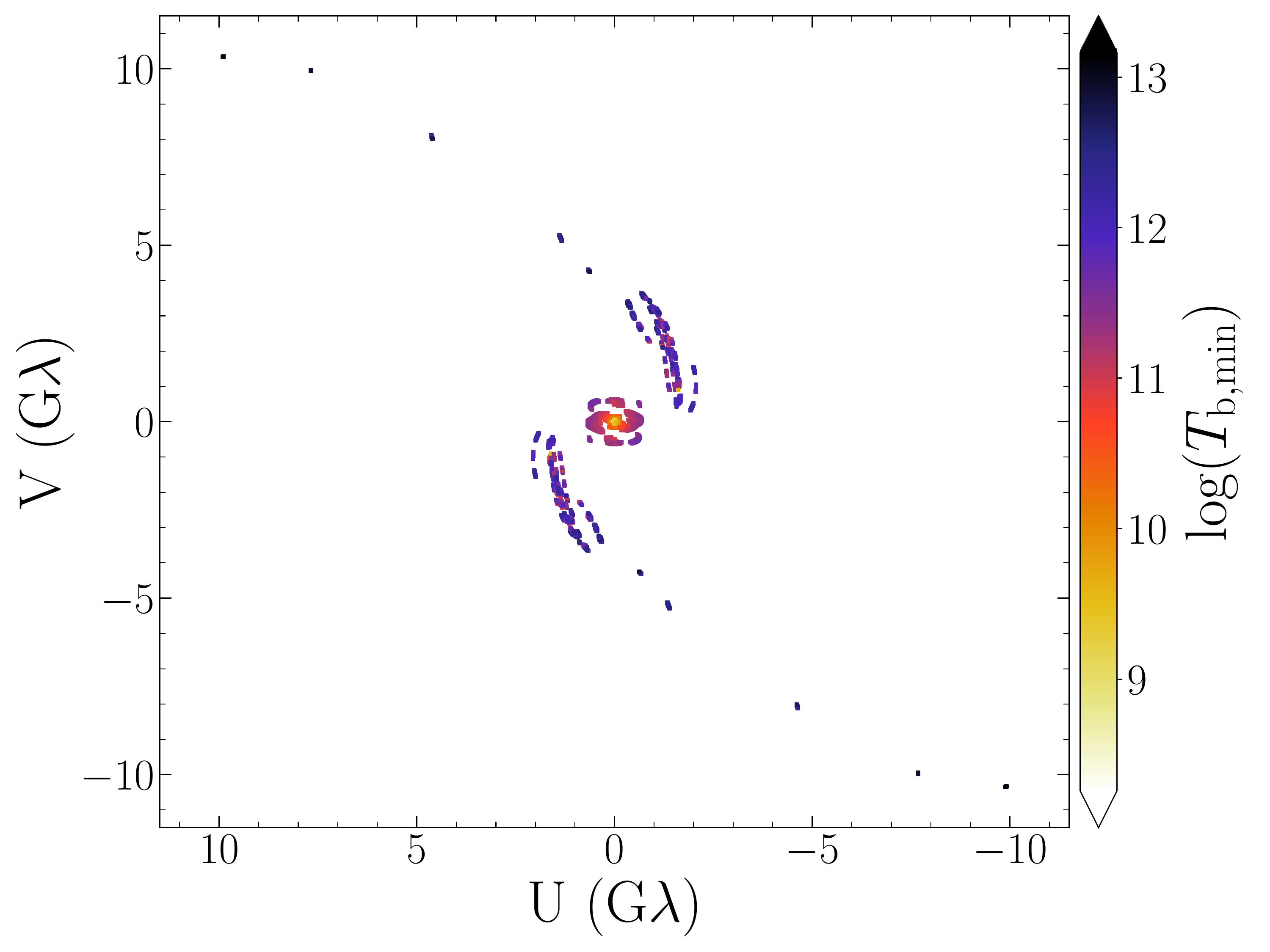}
\caption{Fourier coverage of the fringe-fitted interferometric visibilities of \oj, observed by \ra between March and April 2014 at 22\,GHz. The color range designates the lower limit of the observed brightness temperature (see Sec.~\ref{sec:tb}).
}
\label{fig:uvrad_tb}
\end{figure}

\begin{figure}
\centering
\includegraphics[width=\columnwidth]{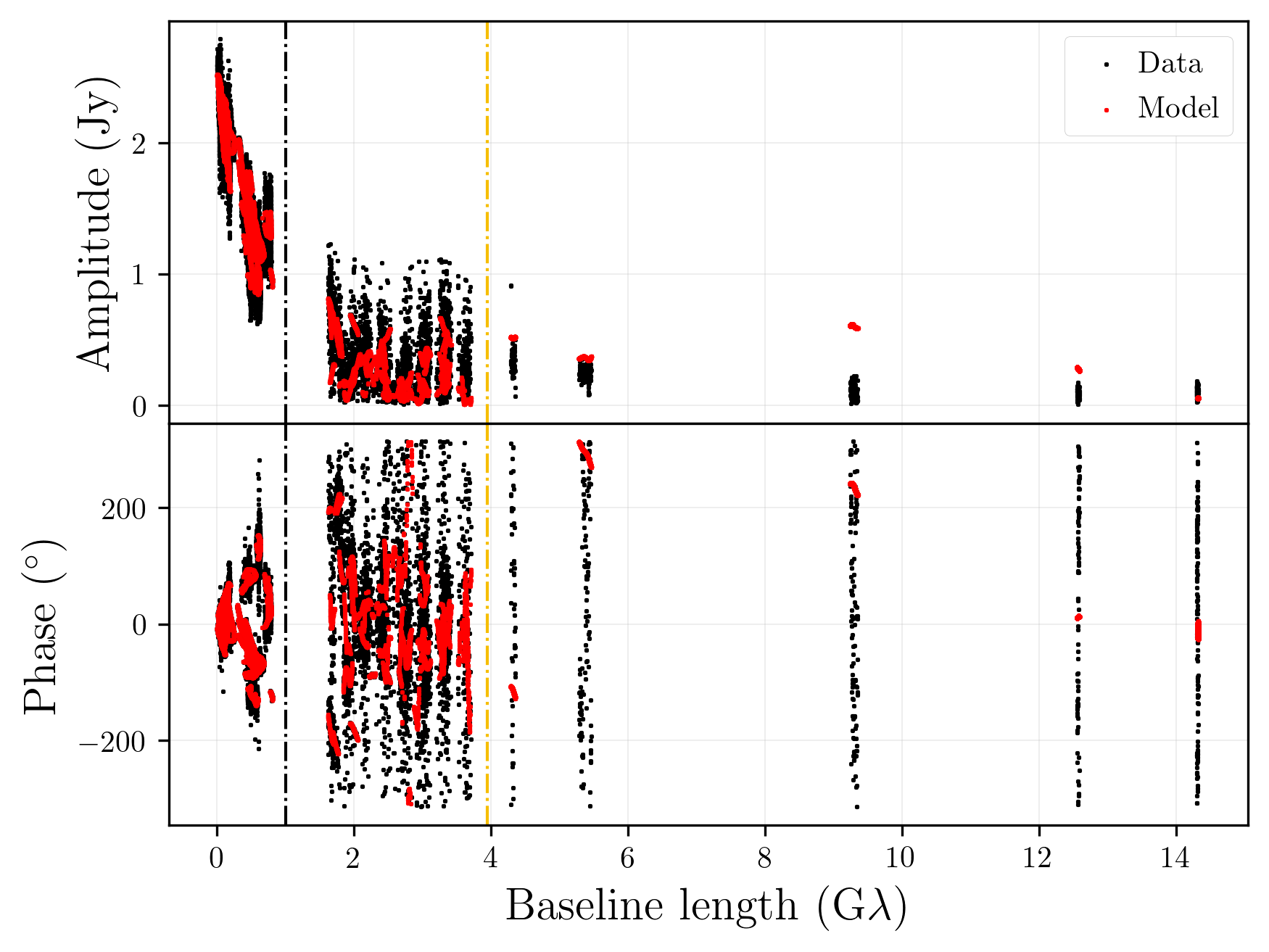}
\caption{Self-calibrated visibility amplitudes and phases as a function of uv-distance of the \ra observations of \oj on March-April 2014 at 22\,GHz. Over-plotted in red is the fit to these data of the CLEAN model obtained from the hybrid mapping using only the data collected during the perigee session, with fringe detections up to a projected baseline length of 3.9 Earth diameters (yellow dot-dashed line). Detections at larger baseline spacing correspond to the long-baseline snapshots. The black dot-dashed line separates the ground-only baselines from those obtained to the SRT.}
\label{fig:RA-radplot}
\end{figure}

Fringes to the SRT were found throughout the whole duration of the perigee imaging session, up to projected baselines of 3.9 Earth diameters in length. The group delay difference between the two polarizations was corrected using AIPS's task RLDLY, and a final complex bandpass function was solved for the receivers before averaging the fringe-fitted data in frequency across each IF and exporting for subsequent imaging.

Fringe fitting, complex bandpass calibration, and \textit{a-priori} amplitude calibration of the long-baseline snapshot survey data were performed using \texttt{PIMA} software \citep{2011AJ....142...35P} as part of the pipeline data processing of the \ra AGN survey \citep{2020AdSpR..65..705K}.

The resulting coverage of the fringe-fitted data in the Fourier domain for the perigee imaging session and the long-baseline snapshots is shown in Fig.~\ref{fig:uvrad_tb}.

Imaging was performed using the {\tt DIFMAP} software \citep{Shepherd:1997wv} following the standard CLEAN \citep{1974A&AS...15..417H} hybrid imaging and self-calibration procedure. Only data from the perigee session (see Fig.~\ref{fig:RA-radplot}) were used for the imaging of the \ra observations. Data from the long-baseline snapshots were excluded based on two arguments. First, no visibility closure quantities that could constrain the imaging and self-calibration procedure were obtained during the long-baseline snapshots, for which only single-baseline fringe detection were obtained. Secondly, the VLBI aperture synthesis technique is based on the assumption that the source remains stationary during the whole integration period being considered. This is no longer valid when considering together the perigee imaging session and the long-baseline snapshots, conducted during different orbits of the SRT (weeks apart), and coincident with a high-activity period in the source, as discussed in Sec.~\ref{sec:single-dish}. This is particularly relevant for the extremely small spatial scales probed during the long-baseline snapshots, with fringe detection spacing between 4.6 $D_\earth$ and 15.1 $D_\earth$. Although the long-baseline snapshot data were not used for the imaging of the \ra data, they provide very valuable information regarding the brightness temperature at the smallest spatial scales, which is analyzed in Sec.~\ref{sec:tb}.

\begin{figure*}
\includegraphics[width=\textwidth]{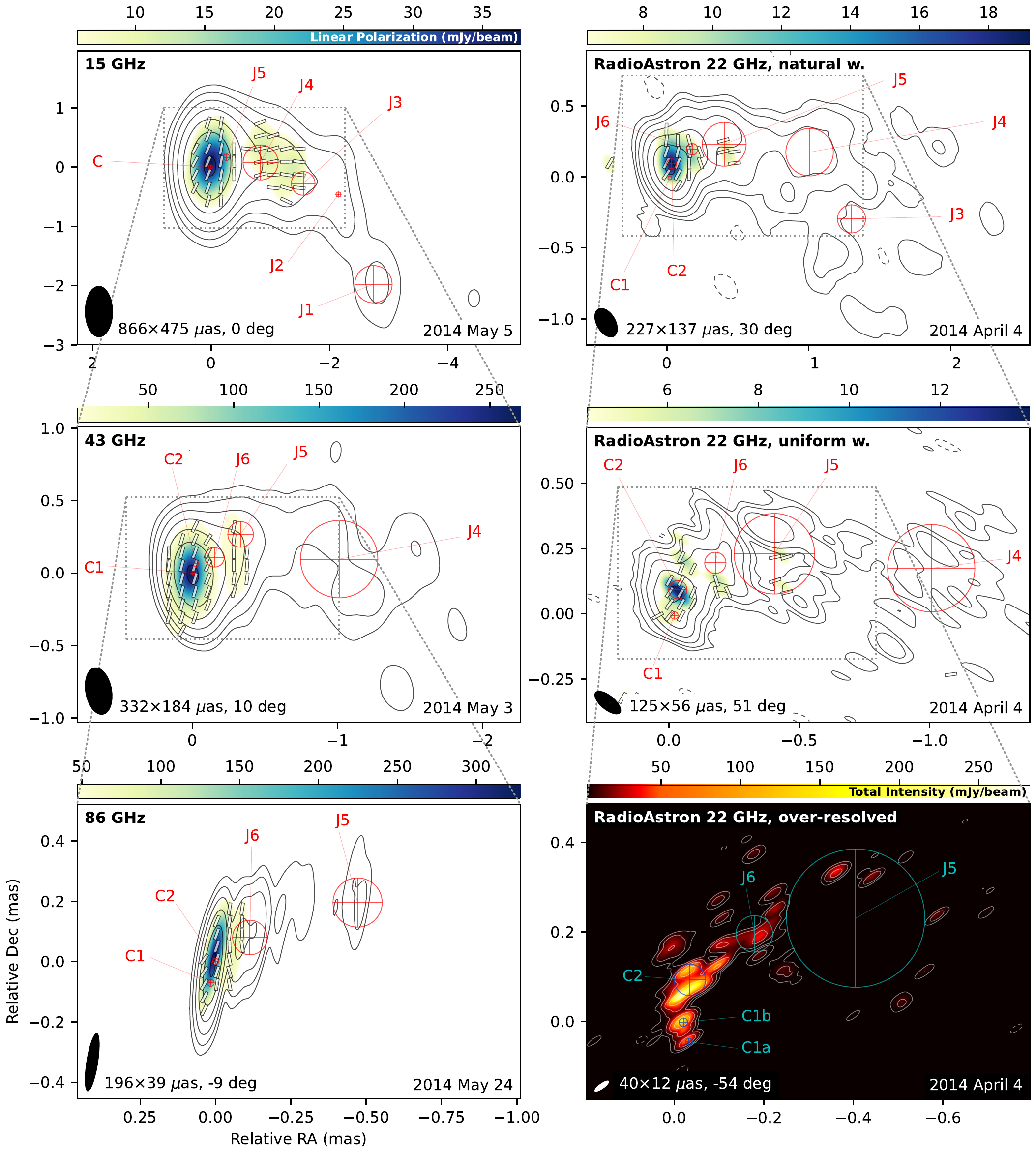}
\caption{From top to bottom, left column: 15, 43, and 86 GHz polarimetric images obtained in May 2014. Right column: \ra 22\,GHz polarimetric space VLBI images obtained with natural and uniform weighting in April 2014. The \emph{over-resolved} total intensity \ra image at the nominal resolution corresponding to the maximum projected baseline fringe detection during the long-baseline snapshots is also shown for reference. Total intensity contours, EVPAs, and model-fit components are over-plotted. The lowest contours are $\pm10$ and $\pm9$ times the rms level (see Tab.~\ref{table:data}) for the 86\,GHz and \ra over-resolved images, respectively, and $\pm7$ for the remaining images, with successive contours in factors of 3 up to 90$\%$ of the total flux density peak. Synthesized beams and their parameters are located in the bottom left corner of each map.}
\label{Fig:all}
\end{figure*}

Perigee-imaging self-calibrated Stokes I visibility amplitudes and phases as a function of Fourier spacing (uv-distance), and CLEAN model fit to these data are shown in Fig.~\ref{fig:RA-radplot}, where we also plot the long-baseline complex visibilities for comparison (see Sec.~\ref{sec:tb} for further discussion). Space VLBI fringes to the SRT extend the projected baseline spacing during the perigee imaging session up to $\sim3.9 D_\earth$, increasing accordingly the angular resolution with respect to that provided by ground-based arrays up to $\sim 56$~$\mu$as for uniform visibility weighting. \ra images of \oj during our April 2014 observations are shown in the right panels of Fig.~\ref{Fig:all}. In this figure we also show for reference the over-resolved image obtained by down-weighting the short baselines and using a Gaussian beam with full width half maximum equal to the nominal resolution of $\sim 12$~$\mu$as corresponding to the longest projected baseline detection of $\sim$15.1 $D_\earth$ obtained during the long-baseline snapshots, although we stress that these data were not used during imaging process.

Calibration of the instrumental polarization and absolute orientation of the polarization vectors is discussed in Appendix \ref{Sec:RA_pol_cal}.

\subsection{Complementary ground VLBI observations at 15, 43 and 86\,GHz}

We complement the analysis of the space VLBI observations of \oj at 22\,GHz with data from three ground-based interferometric observations performed at 15, 43, and 8\,GHz within less than two months from the \ra observations (see Tab.~\ref{table:obs}). At 15\,GHz, a single epoch of fully calibrated visibilities is provided via the publicly available database of the long-term monitoring MOJAVE program\footnote{https://www.physics.purdue.edu/MOJAVE/} \citep{2009AJ....138.1874L} (Principal investigator: J. L. Richards, project code S6407B). The dataset at 43\,GHz is part of the VLBA-BU-BLAZAR program\footnote{https://www.bu.edu/blazars/VLBAproject.html}, which includes monthly observations of 38 radio and $\gamma$-ray bright AGNs. The 43\,GHz visibilities data available on-line are fully self calibrated, following the analysis described in \cite{2017ApJ...846...98J}. 

The 86\,GHz data were obtained from an observation made on 25 May with the GMVA. The duration of each scan was eight minutes. The data were recorded at 2\,Gbps rate (512\,MHz bandwidth) with 2\,bit digitization, apart from for Plateau de Bure observatory, which recorded at 1\,Gbps rate mode and Yebes telescope that was equipped with only LCP receiver. During the correlation procedure, a polyphase filter band technology was used, segmented data into 16 IFs of 32\,MHz bandwidth (8 physical IFs) per polarisation. The data were correlated with the {\tt DiFX} correlator \citep{Deller_2007} at the Max-Planck-Institut f\"ur Radioastronomie in Bonn, Germany. 
 
Data reduction of the GMVA observations was performed using NRAO's {\tt AIPS} software. We first applied the parallactic angle correction, followed by the determination of the inter-band phase and delay offsets between the intermediate frequencies, and the phase alignment across the observing band (known also as manual phase calibration). Next, a global fringe fitting was performed, correcting for the residual delays and phases with respect to a chosen reference antenna. The visibility amplitudes were calibrated, considering the contribution of atmospheric opacity effects based on the measured system temperatures and the gain-elevation curves of each telescope. Finally, the \textit{a-priori} calibrated data were exported to {\tt DIFMAP} for subsequent imaging following the usual CLEAN and self-calibration procedure. The absolute EVPA calibration, as well as the leakage calibration method, is described in Appendix \ref{Sec:Polarization_calib}.

\begin{deluxetable*}{ccccccc}[h]
\tablecaption{Image Parameters.\label{table:data}}
\tablewidth{0pt}
\tablehead{
\colhead{Frequency}  &  \colhead{S$_\mathrm{peak}$} & \colhead{S$_\mathrm{total}$} & \colhead{S$_\mathrm{rms}$} & \colhead{P$_\mathrm{peak}$} & \colhead{P$_\mathrm{total}$} & \colhead{P$_\mathrm{rms}$} \\
\colhead{(GHz)} & \colhead{(mJy/beam)} & \colhead{(mJy)} & \colhead{(mJy/beam)} & \colhead{(mJy/beam)} & \colhead{(mJy/beam)} & \colhead{(mJy/beam)} \\
\colhead{(1)} & \colhead{(2)} & \colhead{(3)} & \colhead{(4)} & \colhead{(5)} & \colhead{(6)} & \colhead{(7)}
}
\startdata
15 &    3281 & 3975 & 7 & 40 & 50 & 1\\
22 &    1184 & 2380 & 7 & 20 & 30 & 93 \\
22$^\mathrm{a}$ & 1278 & 2563 & 2 & 20 & 30 & 15 \\
22$^\mathrm{b}$ & 300 & 2530 & 5 & - & - & - \\
43 &   4510 & 5503 & 15 & 260 & 280 & 16  \\
86 &   3460 & 5051 & 13 & 330 & 460 & 30 \\
\enddata
\tablecomments{$^\mathrm{a}$: All images have been obtained using uniform weighting except for this one, for which natural weighting was used. $^\mathrm{b}$: Over-resolved image of \oj.  Columns from right to left: (1) Observing frequency, (2) Peak flux density, (3) Total flux density, (4) Noise level, (5) Peak of linearly polarized flux density, (6) Total linearly polarized flux density, (7) Noise level in the polarization image.}
\end{deluxetable*}

\subsection{Visibility model fitting}
\label{sec:modelfit}

We modeled the observed brightness distribution in the jet of OJ\,287 with two-dimensional Gaussian components in the visibility plane using the {\tt DIFMAP} package and following the procedure described in \cite{2020A&A...634A.112T}. The imaging and modeling of the multi-frequency VLBI data allowed us to characterize the central compact region of the jet down to parsec-scale. The cross-identification of the individual components across epochs and frequencies was based on consistency of their positions, flux densities and sizes. 
The uncertainties of the component parameters are estimated using the measured root mean square (rms) noise in the respective images and the SNR of detection of individual components \citep[see][for details]{Fomalont:1999uc,Schinzel:2012gm,Poetzl:2021fp}. All of the modelfit components are described in Tab.~\ref{table:knot_parameters} and displayed in Fig.~\ref{Fig:all}.

\begin{deluxetable*}{ccccccccc}
\tablecaption{Model-fitting parameters of OJ~287 during April-May 2014. \label{table:knot_parameters}}
\setlength{\tabcolsep}{2mm}
\tablewidth{0pt}
\tablehead{
\colhead{Knot} & Sub-knot & Sub-knot & \colhead{Freq.} &  \colhead{S} & \colhead{r} & \colhead{PA} & \colhead{FWHM} & \colhead{$T_{b}$} \\
\colhead{Name} & \colhead{Name} & \colhead{Name} & \colhead{(GHz)} & \colhead{(mJy)} & \colhead{($\mu$as)} & \colhead{($^{\circ}$)} & \colhead{($\mu$as)} & \colhead{($10^{10}$ K)} \\
\colhead{(1)} & \colhead{(2)} & \colhead{(3)} & \colhead{(4)} & \colhead{(5)} & \colhead{(6)} & \colhead{(7)} & \colhead{(8)} & \colhead{(9)} 
}
\startdata
C  & ... & ... & 15 & 3100 $\pm$ 220 &    0 $\pm$  45 & ...           &  65 $\pm$    5 &  520 $\pm$  93 \\
   &  C1 & ... & 22 &  520 $\pm$ 110 &    0 $\pm$  20 & ...           &  29 $\pm$   10 &  205 $\pm$ 130 \\
   &  C1 & ... & 43 & 3170 $\pm$ 370 &    0 $\pm$  30 & ...           &       $<$30    &       $>$270   \\
   &  C1 & ... & 86 & 1050 $\pm$ 320 &    0 $\pm$  30 & ...           &  23 $\pm$   10 &   43 $\pm$  34 \\
   &     & C1a & 22 &  120 $\pm$  60 &    0 $\pm$  15 & ...           &  20 $\pm$   15 &   99 $\pm$ 180 \\
   &     & C1b & 22 &  440 $\pm$ 100 &   40 $\pm$  10 &   13 $\pm$ 10 &  21 $\pm$   15 &  330 $\pm$ 535 \\
   &  C2 & ... & 22 & 1170 $\pm$ 160 &  100 $\pm$  15 &$-$12 $\pm$  9 &  70 $\pm$   40 &   79 $\pm$  92 \\
   &  C2 & ... & 43 & 1530 $\pm$ 260 &   70 $\pm$  40 &$-$18 $\pm$ 31 &       $<$45    &        $>$63   \\
   &  C2 & ... & 86 & 2930 $\pm$ 510 &    90 $\pm$  15 & $-$8 $\pm$ 45 &  20 $\pm$    5 &  160 $\pm$  90 \\
\hline
J6 & ... & ... & 22 &  360 $\pm$  90 &  260 $\pm$  30 &$-$38 $\pm$  6 &  80 $\pm$   55 &   19 $\pm$  26 \\
J6 & ... & ... & 43 &  710 $\pm$ 180 &  190 $\pm$  70 &$-$45 $\pm$ 20 & 130 $\pm$   50 &    4 $\pm$   3 \\
J6 & ... & ... & 86 &  890 $\pm$ 290 &  200 $\pm$  45 &$-$36 $\pm$ 13 & 114 $\pm$  150 &    1 $\pm$   4 \\
\hline
J5 & ... & ... & 15 &  500 $\pm$  90 &  320 $\pm$ 115 &$-$58 $\pm$ 20 &      $<$140    &        $>$18   \\
J5 & ... & ... & 22 &  220 $\pm$  70 &  450 $\pm$  55 &$-$59 $\pm$  7 & 150 $\pm$  190 &    3 $\pm$   8 \\
J5 & ... & ... & 43 &  210 $\pm$ 100 &  430 $\pm$ 150 &$-$51 $\pm$ 19 & 170 $\pm$  100 &  0.6 $\pm$ 0.9 \\
J5 & ... & ... & 86 &  120 $\pm$ 130 &  470 $\pm$ 205 &$-$61 $\pm$ 24 & 163 $\pm$  415 &  0.1 $\pm$ 0.5 \\
\hline
J4 & ... & ... & 15 &  230 $\pm$  60 &  860 $\pm$ 230 &$-$85 $\pm$ 15 & 590 $\pm$  330 &  0.5 $\pm$ 0.6 \\
J4 & ... & ... & 22 &  140 $\pm$  60 & 1000 $\pm$ 200 &$-$79 $\pm$ 11 & 460 $\pm$ 1840 &  0.2 $\pm$   2 \\
J4 & ... & ... & 43 &  150 $\pm$  90 & 1020 $\pm$ 345 &$-$82 $\pm$ 19 & 510 $\pm$  810 & 0.05 $\pm$ 0.2 \\
\hline
J3 & ... & ... & 15 &   80 $\pm$  40 & 1590 $\pm$ 355 &$-$100 $\pm$ 13 & 400 $\pm$  205 &  0.4 $\pm$ 0.4 \\
J3 & ... & ... & 22 &   50 $\pm$  40 & 1330 $\pm$ 195 &$-$109 $\pm$  8 & 230 $\pm$  540 &  0.3 $\pm$   2 \\ 
\hline
J2 & ... & ... & 15 &   30 $\pm$  30 & 2210 $\pm$ 610 &$-$102 $\pm$ 15 &      $<$550    &      $>$0.07   \\
J2 & ... & ... & 22 &   40 $\pm$  40 & 1920 $\pm$ 295 &$-$127 $\pm$  9 & 306 $\pm$  990 &  0.1 $\pm$ 0.9 \\
\hline
J1 & ... & ... & 15 &   50 $\pm$  30 & 3400 $\pm$ 560 &$-$126 $\pm$  9 & 630 $\pm$  490 & 0.09 $\pm$ 0.2 \\ 
\hline
\enddata
\tablecomments{Columns from left to right: (1-3) Component/sub-component ID, (4) Observing frequency, (5) Flux density, (6) Radial distance from the core, (7) Position angle, (8) Component size, (9) Brightness Temperature (source frame).}
\end{deluxetable*}


\section{Results}
\label{sec:results}

\begin{deluxetable}{cccc}
\tablecaption{Polarization properties of the modelfit components of OJ~287 during April-May 2014. The components are presented in order of increasing distance from the reference point of each image.
\label{table:pol_data}}
\tablewidth{0pt}
\tablehead{\colhead{Knot} & \colhead{Freq.} &  \colhead{m} & \colhead{$\chi$} \\
\colhead{Name} & \colhead{(GHz)} & \colhead{\%} & \colhead{($^{\circ}$)} \\
\colhead{(1)} & \colhead{(2)} & \colhead{(3)} & \colhead{(4)}
}
\startdata
P$_\mathrm{C}$ & 15  & 1.2 $\pm$ 0.4   & $-$22.5 $\pm$ 0.7 \\
\hline
\multirow{3}{*}{P$_\mathrm{C1}$}  & 22 & 1.5 $\pm$ 0.5  & $-$35.4 $\pm$ 5.6 \\
   & 43  & 8.2 $\pm$ 0.8   & $-$16.2 $\pm$ 5.0 \\
   & 86  & 20.9 $\pm$ 3.3   & $-$23.4 $\pm$ 7.0 \\   
   \hline
\multirow{3}{*}{P$_\mathrm{C2}$}  & 22 & 1.0 $\pm$ 0.6  & $-$24.18 $\pm$ 5.6 \\   
   & 43  & 15.9 $\pm$ 1.6   & $-$17.5 $\pm$ 5.0 \\
   & 86  & 10.4 $\pm$ 1.3   & $-$20.5 $\pm$ 7.0 \\ 
   \hline
\multirow{3}{*}{P$_\mathrm{J6}$}  & 22 & 1.1 $\pm$ 0.6  & 17.6 $\pm$ 5.6 \\   
   & 43  & 6.3 $\pm$ 0.6   & $-$20.7 $\pm$ 5.0 \\
   & 86  & 5.8 $\pm$ 4.0   &  8.9 $\pm$ 7.0 \\ 
   \hline
\multirow{2}{*}{P$_\mathrm{J5}$}  & 15 & 4.8 $\pm$ 1.1  & $-$8.9 $\pm$ 1.3 \\ 
   & 43  & 5.8 $\pm$ 0.6   & $-$7.6 $\pm$ 5.0 \\
      \hline
P$_\mathrm{J4}$ & 15  & 3.0 $\pm$ 0.5   & $-$83.9 $\pm$ 1.0 \\
   \hline
P$_\mathrm{J3}$ & 15  & 7.5 $\pm$ 1.6   & $-$79.5 $\pm$ 0.8 \\
\enddata
\tablecomments{The columns from left to right: (1) Component ID, (2) Observing frequency, (3) the degree of linear polarization, and (4) the EVPA.}
\end{deluxetable}

\subsection{Space and millimeter VLBI images of \oj}
\label{Sec:Imaging}

\subsubsection{Total Intensity}
\label{subsec:total_intensity}

Figure~\ref{Fig:all} shows the ground array (left panels) and \ra space VLBI images (right panels) of \oj taken between April and May 2014, providing a detailed view of the jet at different spatial scales.

The 15\,GHz image (left upper panel in Fig.~\ref{Fig:all}) shows a bright VLBI core and a one-sided jet that extends towards the west, followed by a strong jet bending by about 55$^{\circ}$ towards the south-west. The core region is modeled by a single Gaussian component labeled as C (upper left frame in Fig.~\ref{Fig:all}), considered as the positional reference for the remaining modelfit components. At higher frequencies, this region is resolved into two sub-components C1 and C2 of the core and an extra jet component J6 (see Fig.~\ref{Fig:all}). 

At 22, 43 and 86\,GHz images the modelfit component C1 is considered as the reference point. In the over-resolved \ra image, the knot C1 can be further resolved into two sub-components C1a and C1b (bottom right frame in Fig.~\ref{Fig:all}). We note that the sum of the flux densities of C1a and C1b components approximately adds to that of component C1 (see Tab.~\ref{table:knot_parameters}).

The progressively higher angular resolution obtained with the VLBA-BU-BLAZAR 43\,GHz image (middle left panel in Fig.~\ref{Fig:all}), and GMVA 86\,GHz image (bottom left panel) allows us to map the innermost regions of the VLBI core area up to an angular resolution of about 40\,$\mu$as. This reveals that the jet bending observed at 15\,GHz continues as we probe deeper into the VLBI core, from the west jet direction observed at 15\,GHz to the north-west jet direction visible at 86\,GHz. The jet curvature at these spatial scales has also been reported by \cite{2017A&A...597A..80H}, and is clearly imprinted here as well in the position angle of the model fitted components listed in Tab.~\ref{table:knot_parameters}: the position angle of the innermost component rotates from $-58.4^{\circ}\pm0.4^{\circ}$ for component J5 at 15\,GHz, to $-18.4^{\circ}\pm1.4^{\circ}$ for C2 at 43\,GHz, and finally $-8.1^{\circ}\pm1.1^{\circ}$ for component C2 at 86\,GHz.

Our space VLBI images of \oj (right panel in Fig.~\ref{Fig:all}) confirm that this progressive jet bending with increasing angular resolution continues up to the smallest scales probed by \ra. At these extremely high angular resolutions the core area can be resolved into two distinct components C1a and C1b. Assuming that the upstream end of the jet characterized by component C1a corresponds to the VLBI core, the innermost jet depicted by the \ra images shows a counter-clock wise rotation from $-11.7^{\circ}\pm1.0^{\circ}$ for C2 to $13.1^{\circ}\pm0.4^{\circ}$ for component C1b, as we probe deeper into the upstream jet.

Downstream from the core region, the jet structure is well represented by up to six Gaussian components, labeled J1 through J6 (see Fig.~\ref{Fig:all}). Except for the outermost component J1, all other jet components are cross-identified at multiple frequencies.

\begin{figure*}
\centering
\includegraphics[width=0.7\textwidth]{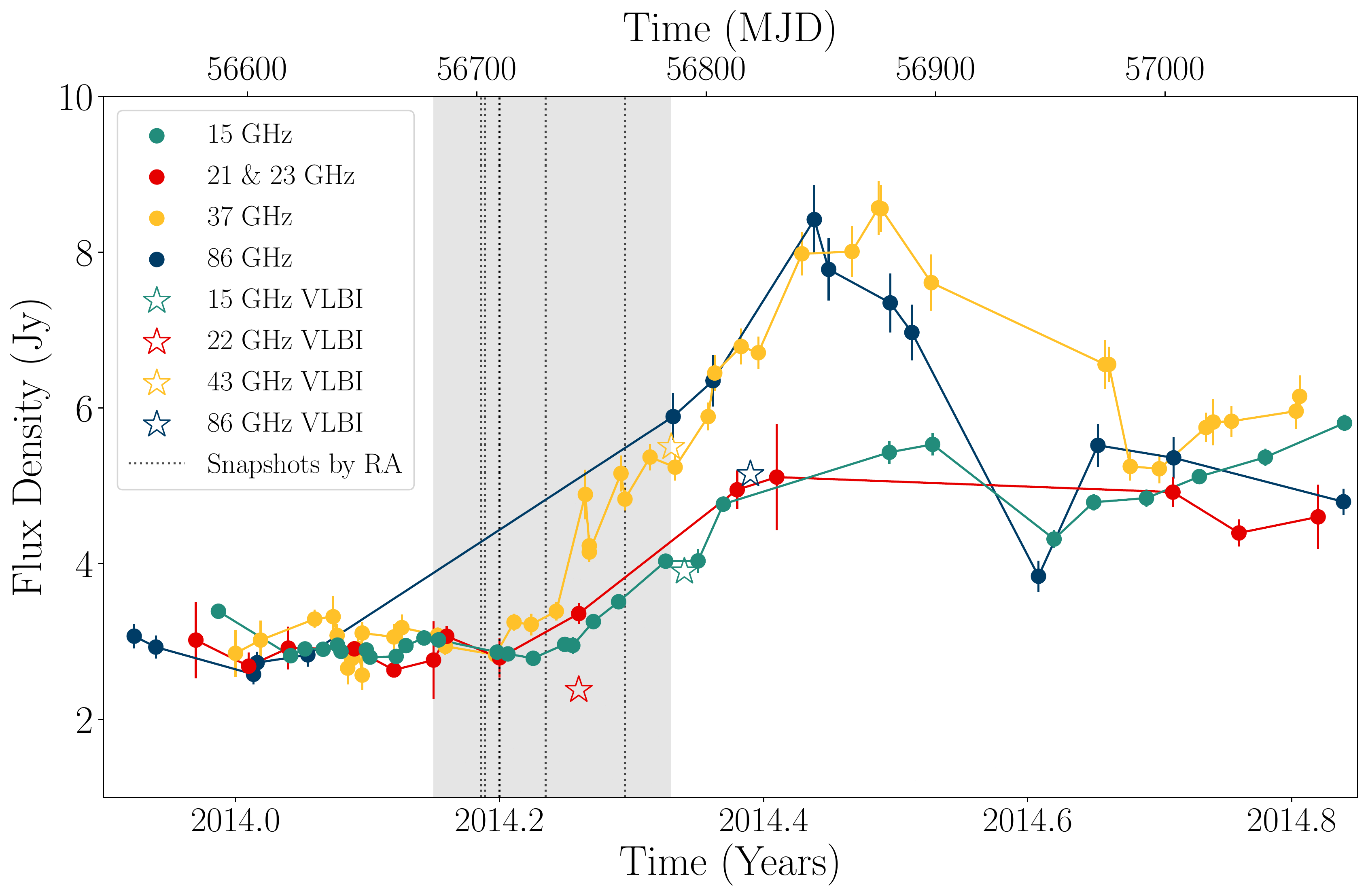}
\caption{Single-dish (circles) flux density multifrequency measurements. During our observing interval, a prominent flaring event took place, initiated by increased $\gamma$-ray activity, displayed by the shadowed, vertical area. The dotted vertical lines designate the long-baseline \ra snapshot sessions, whereas the open stars the VLBI observations.}
\label{fig:single_dish}
\end{figure*}

We note here that one can try to decompose the 15\,GHz flux density of the component C (3.1\,Jy) into contributions from C1 and C2 by calculating their spectral indices between 22\,GHz and 43\,GHz and using them to estimate their respective flux densities at 15\,GHz. This procedure yields the following estimates: $S_\mathrm{15GHz,C1}=0.2$\,Jy and $S_\mathrm{15GHz,C2}\approx 1.0$\,Jy. Taken together, they sum up to 1.2\,Jy, which is about 1.9\,Jy less than the flux density actually measured at 15\,GHz for the component C. This discrepancy most likely results from the fact that the 22\,GHz \ra observations were made 29 days before the 43\,GHz observation and 31 days before (see Fig. \ref{fig:single_dish} and Sec.~\ref{sec:spix}), where \oj was in a lower flux density state. As \oj was undergoing the rising stage of a flare during this period (see Sec.~\ref{sec:single-dish}), the discrepancy of 1.9\,Jy between the estimated and measured flux density at 15\,GHz suggests that the 22\,GHz flux density of one or both components (C1 and C2) blended at 15\,GHz into the single component C has increased during this period. 

The fitted flux densities and sizes of the modelfit components were used to calculate also the brightness temperature of each VLBI knot in the rest frame of the source  \citep[e.g.,][]{2012A&A...544A..34P}:
\begin{equation}
    T_{b} = 1.22 \times 10^{12} \frac{S}{\theta_\mathrm{obs}^{2} \nu^{2}} \left( 1+z \right) ~~\left( \mathrm{K} \right)\, ,
    \label{eq:Tb}
\end{equation}
where $S$ is the component flux density in Jy, $\theta_\mathrm{obs}$ is the size of the emitting region in mas, $\nu$ is the observing frequency in GHz, and $z$ is the source redshift.  For unresolved modelfit components, we set
$\theta = \theta_\mathrm{min}$, where $\theta_\mathrm{min}$ is the resolution limit \citep{2005astro.ph..3225L} obtained for the respective component, and consider the resulting estimate of $T_\mathrm{b}$ as a lower limit. The brightness temperatures estimated using this procedure are given in column 9 of Tab.~\ref{table:knot_parameters}. 

\subsubsection{Polarization}

Linear polarization is detected in all of our images in Fig.~\ref{Fig:all}, with the exception of the over-resolved \ra image. Tab.~\ref{table:pol_data} summarizes the linear polarization properties of the model fitted components, whereas in the Appendix \ref{subsec:polarization} we present the polarization error estimation method that we followed. As it is commonly found in VLBI imaging, there is not a one-to-one correspondence between the components seen in total intensity and linear polarization. Hence, while the model fitting for total intensity is performed in the visibility plane, the two-dimensional distribution of the fractional polarization and the electric vector position angle (EVPA) for each knot are calculated in the image domain, based on the amount of polarized emission inside the area of each component. The polarization structure at 15\,GHz consists of a low polarized (m\,$\sim1.2\,\%$) core with EVPAs in the direction of the innermost jet, and polarized emission in the extended jet between model fitted components J4 and J3. Our characterization of the polarized emission in the jet as components P$_\mathrm{J4}$ and P$_\mathrm{J5}$ (see Tab.~\ref{table:pol_data}) shows an increased degree of polarization with respect to that of the core from $3\,\%$ to $7.5\%$, as expected in the case of lower opacity, and polarization vectors aligned with the local jet direction. As discussed below in Sec.~\ref{subsec:rm}, the observed EVPAs are not severely affected by Faraday rotation effects, from which we can conclude that the VLBI core and jet in \oj are characterized by a dominant toroidal magnetic field component.

The linearly polarized emission at 43 and 86\,GHz shows a progressive increase in degree of polarization with observing frequency in the core, reaching m\,$\sim\,21\%$ at 86\,GHz, which is consistent with a transition from an optically thick core at 15\,GHz to optically thin at 86\,GHz, as discussed in more detail in Sec.~\ref{sec:spix} below, and may be also affected by beam depolarization \citep{1966MNRAS.133...67B}. The emission downstream of the VLBI core area at 43 and 86\,GHz, characterized as components P$_\mathrm{J5}$ and P$_\mathrm{J6}$, shows EVPAs predominantly oriented perpendicular to the local jet direction, and a higher degree of polarization than what is observed further downstream in the jet area corresponding to components P$_\mathrm{J4}$ and P$_\mathrm{J3}$ at 15\,GHz. This suggests that the jet area of P$_\mathrm{J5}$, and maybe the region comprised between components P$_\mathrm{J5}$ and P$_\mathrm{J6}$ as well, may correspond to a recollimation shock that compresses the magnetic field in the direction of the jet, explaining the different EVPAs. However, we should also note that according to numerical simulations \citep[e.g.,][]{2018ApJ...860..121F,2021A&A...650A..61F,2021A&A...650A..60M} the observed net polarization in recollimation shocks can change significantly depending on the underlying magnetic field configuration, viewing angle, and other jet parameters. Alternatively, as mentioned in \cite{sasada18}, the observed polarization may be originated by an oblique shock located where the jet bends to the west.

Alternatively, as suggested by \cite{2018ApJ...860..121F,2021A&A...650A..61F}, EVPA depends on the helical magnetic field pitch angle. Lastly, is it worth mentioning that the jet presents a bend in that location, so it could correspond to an oblique shock \citep{sasada18}.

A higher degree of magnetic field ordering is expected in the shocked plasma, which would also explain the higher degree of polarization in this region \citep{Jorstad_2007,2016A&A...596A..78H}.

The space VLBI \ra images from the perigee imaging (top and middle right panels in Fig.~\ref{Fig:all}) show low degrees of linear polarization in the core area corresponding to components C1 and C2, with values of the order of 1\% (see Tab.~\ref{table:pol_data}), and EVPAs aligned with the local direction of the jet. These low degrees of polarization are expected for optically thick emission, in agreement with what was observed at 15\,GHz. The EVPAs are also in concordance with a predominant toroidal magnetic field, or a helical magnetic field with a large pitch angle. The natural weighted image (top right panel in Fig.~\ref{Fig:all}) shows weakly polarized emission in the area associated with component J6, with EVPAs perpendicular to the local jet direction, providing further support for the possibility that this corresponds to a recollimation shock. The measured degree of polarization for component P$_\mathrm{J6}$ is smaller than that observed at lower frequencies, which suggests that this component may be also affected by opacity effects at 22\,GHz. Polarization information is resolved out in the over-resolved \ra image (bottom right panel in Fig.~\ref{Fig:all}).

\subsection{Flaring activity of \oj during 2014}
\label{sec:single-dish}

\oj showed multi-wavelength flaring activity during April-May 2014, coincident with our VLBI imaging campaign, as shown in the light curves of Fig.~\ref{fig:single_dish}. The event started at high energies, with $\gamma$-ray emission rising on the 25th of February 2014 (shadowed area in Fig.~\ref{fig:single_dish}). For the complete $\gamma$-ray light curve see the VLBA-BU-BLAZAR website\footnote{$\mathrm{http://www.bu.edu/blazars/VLBA\_GLAST/oj287.html}$}. A local optical V band maximum was reported on March 31st \citep{2014ATel.6054....1G}, followed few days later by a flux density increase by almost $\sim$270\% from the flux density level recorded between January to February, as measured by single-dish observations at 21, 23, 37, and 86\,GHz. According to Weaver et al. 2021, (submitted) this flare is associated with the appearance of new features in the \oj jet. The quasi-simultaneous to the VLBI observations light curves were provided by the 40-m telescope of the Owens Valley Radio Observatory (OVRO) at 15\,GHz \citep{2011ApJS..194...29R}, the RATAN-600 radio telescope at 5, 8, 11, and 22\,GHz \citep{1999A&AS..139..545K,2002PASA...19...83K}, the F-Gamma multifrequency monitoring program \citep{2019A&A...626A..60A} at 23\,GHz, the monitoring project of extragalactic radio sources by Mets\"{a}hovi Radio Telescope \citep{1998A&AS..132..305T} at 37\,GHz, and the POLAMI monitoring program \footnote{\url{www.polami.iaa.es/}} at 86\,GHz \citep{2018MNRAS.474.1427A,2018MNRAS.473.1850A}.

We note that the non-simultaneity of VLBI data acquisition under such flaring conditions can introduce a big uncertainty to the spectral index estimation and the Faraday-rotation analysis results, discussed in Sec.~\ref{sec:spix} and Sec.~\ref{subsec:rm}.

\subsection{Spectral analysis}
\label{sec:spix}

The quasi-simultaneous multi-frequency observations in April-May 2014 enable us to analyze the spectral properties of the \oj jet from parsec to sub-parsec scales during its flaring activity. For this analysis, we consider the spectrum, $S(\nu)$, of synchrotron self-absorbed (SSA) emission from a homogeneous spherical region filled with relativistic electron-positron plasma with a power-law energy distribution of emitting particles \citep{2000A&A...361..850T}:

\begin{equation}
 S_{\nu} = S_\mathrm{m} \left( \dfrac{\nu}{\nu_\mathrm{m}} \right)^{\alpha_\mathrm{t}}
 \dfrac{ 1 - \exp[-\tau_\mathrm{m}(\nu / \nu_\mathrm{m})^{\alpha - \alpha_\mathrm{t}}]}{1 - \exp (-\tau_\mathrm{m}) } ~~~~ [\mathrm{Jy}],
 \label{eq:Sm}
\end{equation}
where $S_{\nu}$ is the observed flux density in Jy, $\nu_\mathrm{m}$ is the turnover frequency in GHz, $S_\mathrm{m}$ is the turnover flux density in Jy, $\tau_\mathrm{m} \sim 3/2 \left[ \left( 1-\left(8 \alpha / 3\alpha_\mathrm{t}\right) \right)^{1/2} - 1 \right]$ is the optical depth at $\nu_\mathrm{m}$, and $ \alpha_\mathrm{t}$ and $\alpha$ are the spectral indices of the optically thick and thin parts of the spectrum, respectively (using the $S_{\nu} \propto \nu^{\alpha}$ definition of spectral index).  

\begin{figure*}[ht!]
\centering
\centerline{\includegraphics[align=c,width=0.475\textwidth]{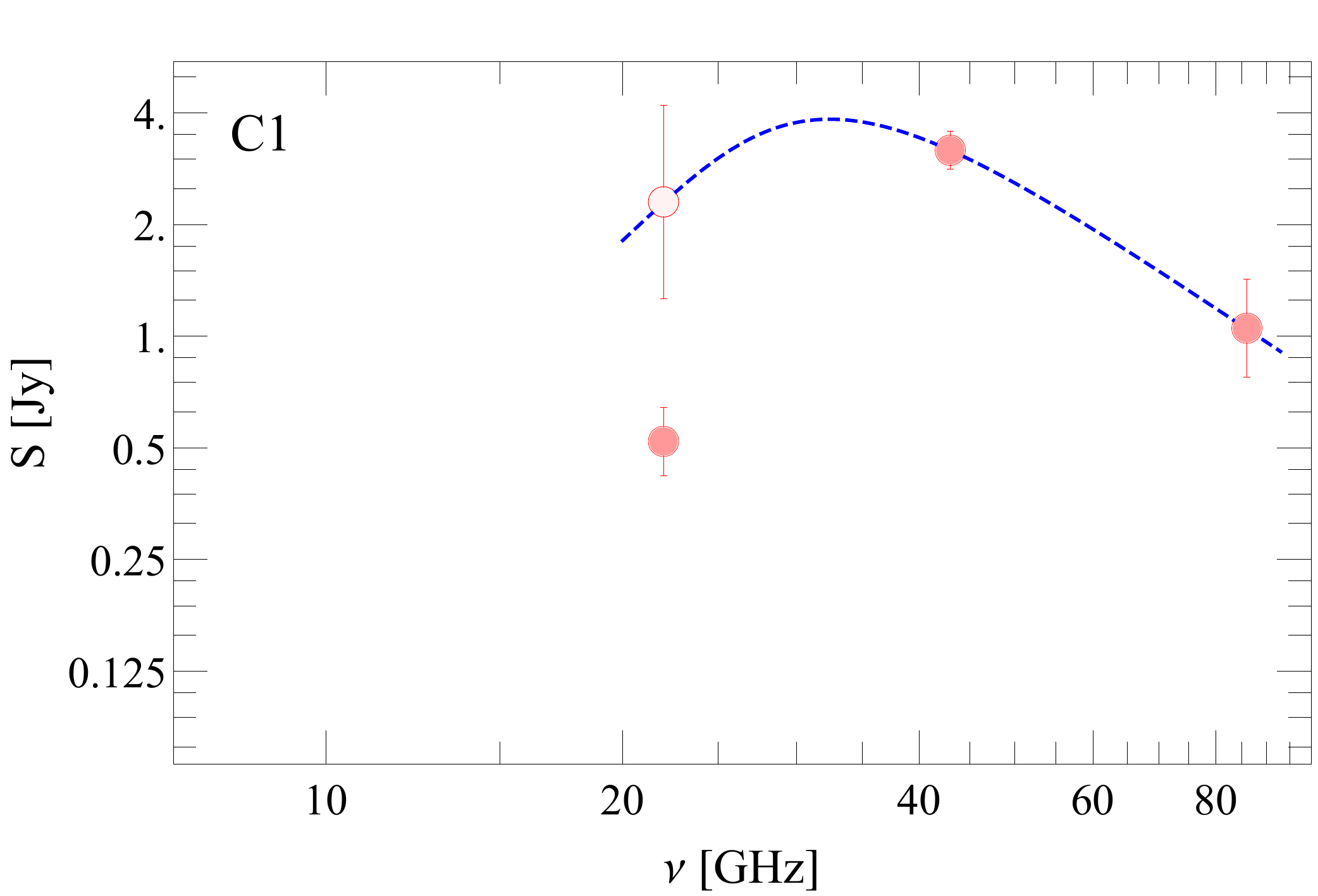}
\includegraphics[align=c,width=0.475\textwidth]{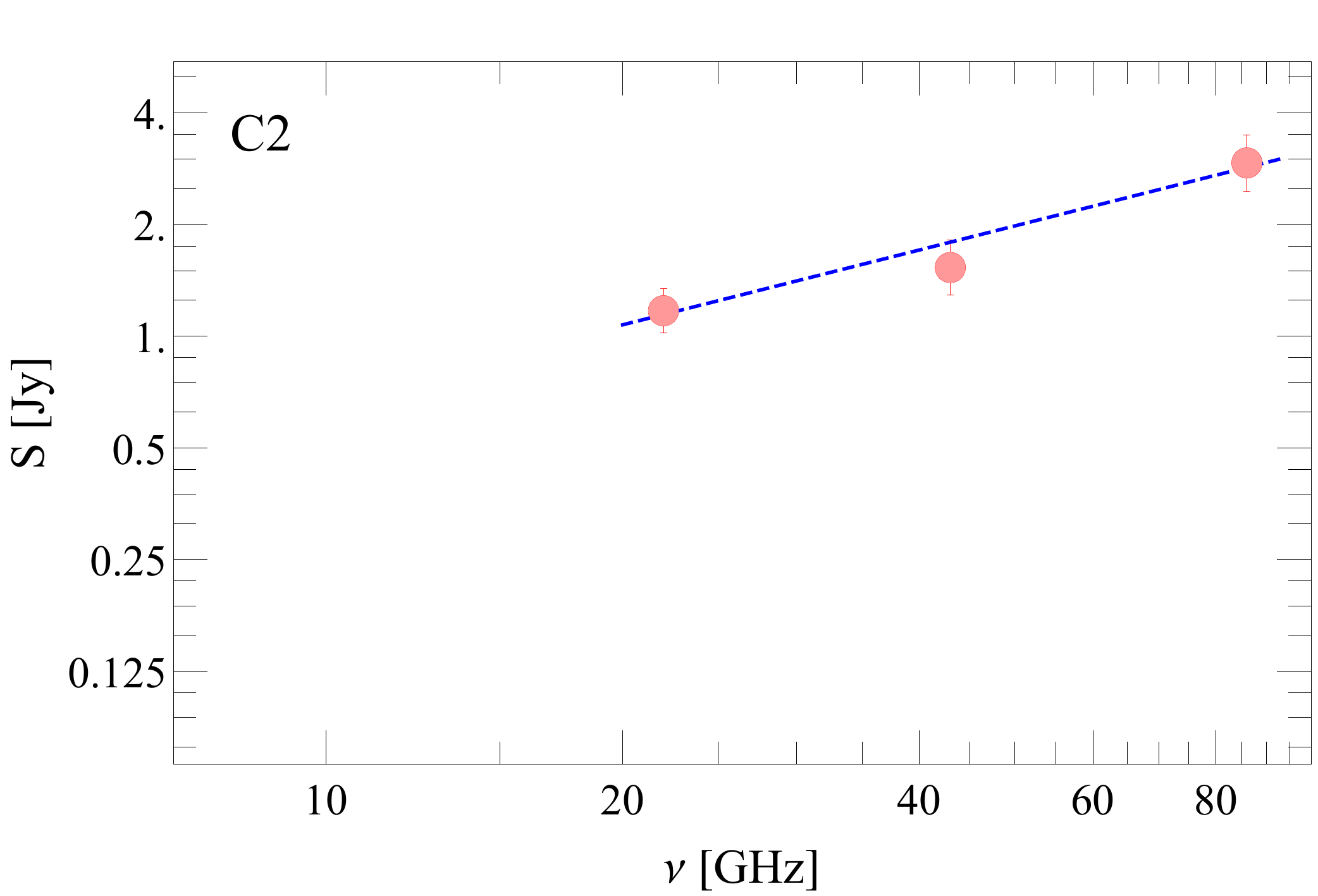}}
\centerline{\includegraphics[align=c,width=0.475\textwidth]{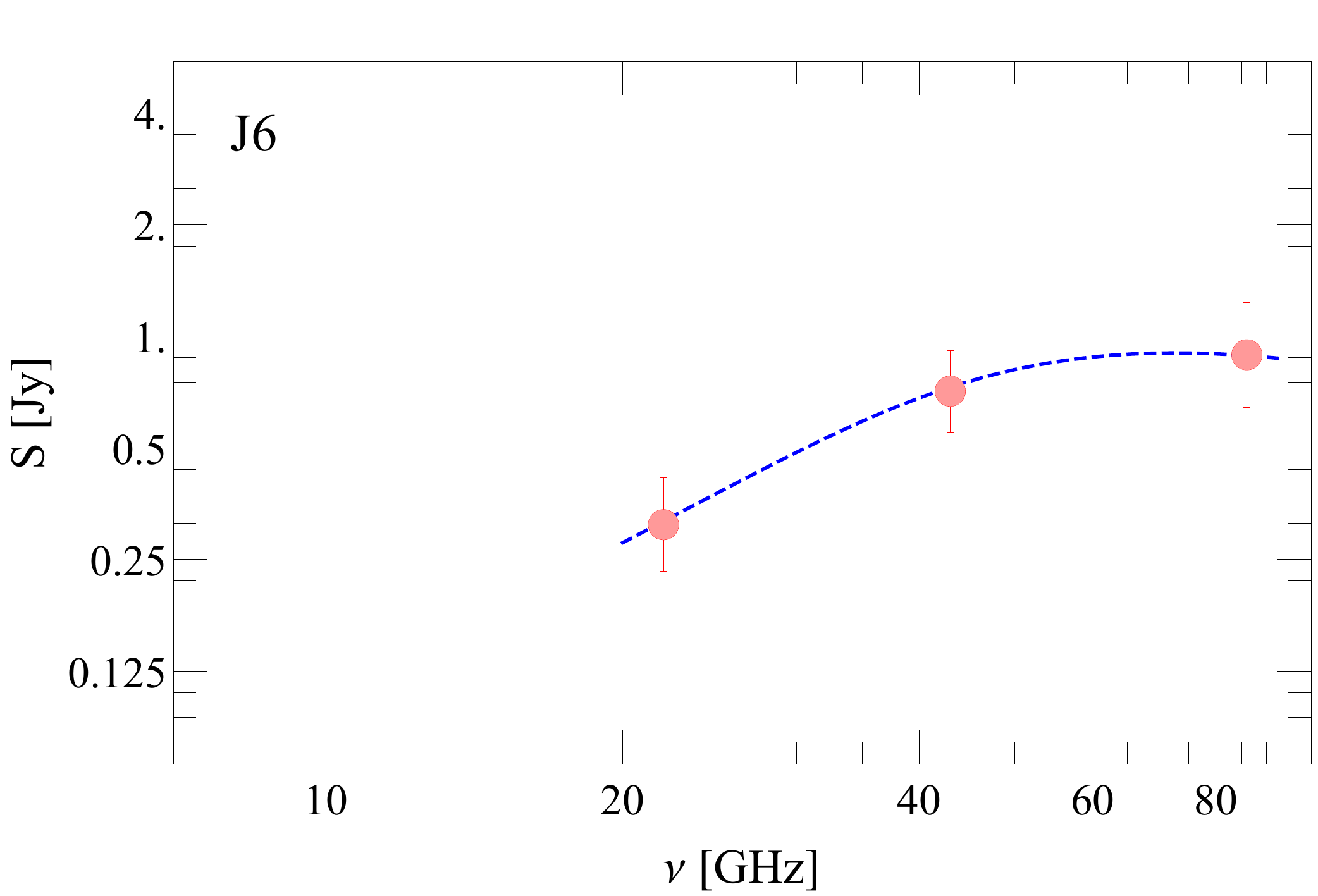}
\includegraphics[align=c,width=0.475\textwidth]{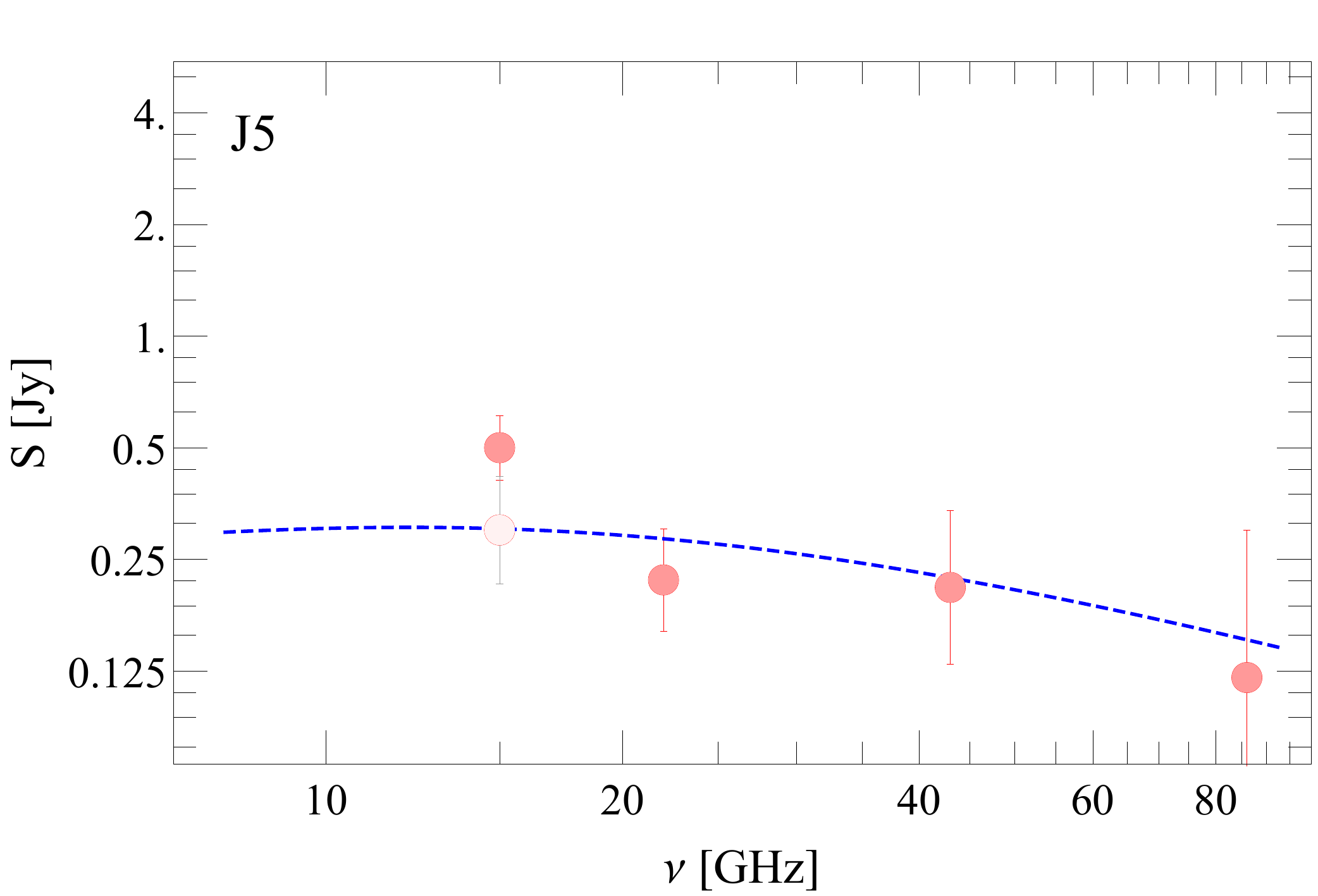}}
\caption{Observed spectra of the jet components C1, C2, J5, and J6 and best fit models to the spectra presented in Tab.~\ref{table:comp_spix}. White circles represent flux densities corrected for the source variability at 22 \,GHz (component C1) and for the blending effect due to the limited resolution at 15\,GHz (component J5).}
\label{fig:comp_spix}
\end{figure*}

In our data, crude estimates of the synchrotron turnover point can be made for four components (C1, C2, J6, and J5) by using the Equation~\ref{eq:Sm} and making an assumption about one of the two spectral slopes. As the 22\,GHz observation with \ra (epoch $t_1=2014.258$, in further discussion) preceded the observations at other frequencies by up to 50 days, we attempt to account for potential effect of source variability by estimating the 22\,GHz flux density at the epoch of May 14 (epoch $t_2=2014.368$, hereafter) situated in the middle of the time period between the VLBI observations at 43\,GHz and 86\,GHz. 

At the epoch $t_1$ of the \ra observation, a total flux density $S_\mathrm{t1}= 2.98\pm0.04$\,Jy was measured at the Effelsberg 100-m telescope. For the epoch $t_2$, we use the Gaussian process predictor with the 21--23\,GHz measurements shown in Fig.~\ref{fig:single_dish} to estimate $S_\mathrm{t2}= 4.84\pm0.07$\,Jy. Hence the 22\,GHz emission of \oj should have increased by about 1.8\,Jy between the two epochs.

To decide on the most plausible distribution of the estimated 1.8\,Jy increase of $S_\mathrm{var,22}$, we inspect the component spectra obtained using the actually measured flux densities (see Fig.~\ref{fig:comp_spix}). This inspection suggests that the 22\,GHz flux densities of the components C1 is the most likely to be affected by the variability between the epochs $t_1$ and $t_2$. Its flux density at the epoch $t_2$ should have then increased to $\approx 2.3$\,Jy. If we use this value to repeat the 15\,GHz flux density decomposition of the core component C discussed in Sec.~\ref{subsec:total_intensity}, we obtain $S_\mathrm{15GHz,C1}\approx 1.9$\,Jy. The respective estimated flux density of the component C then becomes $\approx 3.1$\,Jy, which is in excellent agreement with the actually measured flux density of this component. This agreement further supports the suggestion that the variability observed at 22\,GHz between the epochs $t_1$ and $t_2$ can be ascribed to flux density changes of the component C1 in the core region. We therefore apply this assumption for the spectral fitting described below.

A clear peak in the spectrum is observed only in the component C1, while the component J6 has a rising spectrum, and the component J5 has a falling spectrum. For the component C2, the observed spectrum does not warrant estimating the turnover point, and crude limits on $S_\mathrm{m}$ and $\nu_\mathrm{m}$ can be provided by the component flux density measured at 86\,GHz. If we assume that the turnover flux density of C2 is similar to that of C1, then the turnover frequency of C2 should be $\approx 115$\,GHz. 
For the component J6, the measured flux densities can be used for estimating $S_\mathrm{m}$ and $\nu_\mathrm{m}$ only if an assumption is made about the spectral index $\alpha$ of the optically thin part of the spectrum. We assume $\alpha=-0.7$ for this component. For the component J5, the 15\,GHz flux density is likely to be affected by blending with the component J6 upstream. We estimate this blend to contribute $\approx 0.2$\,Jy to the flux density of J5, and we correct for this blend before fitting the spectrum. The fit is then done with an assumption of $\alpha=-0.8$ derived from the spectral index measured for J5 between 43 and 86 GHz. The resulting overall constraints for the component spectra and the best fit models are presented in Tab.~\ref{table:comp_spix}. The fits are also plotted in Fig.~\ref{fig:comp_spix}.

Results of the spectral fitting indicate an unusual spectral evolution, with the turnover frequency first rapidly rising downstream from the component C1 (with $\nu_m > 86$\,GHz in the component C2) and then falling back. Such a behavior can be explained by relativistic shocks undergoing a transition from Compton- to synchrotron-dominated emission regime \citep{1990psrj.conf..236M}, although reaching viable conclusions on this matter requires taking into account the acceleration of the emitting plasma \citep{1999ApJ...521..509L}. 

\setlength\tabcolsep{2pt}
\begin{deluxetable}{c|cccc}[tb]
\tablecaption{Spectral properties of selected jet components \label{table:comp_spix}}
\tablewidth{0pt}
\tablehead{
Comp. & \colhead{$\alpha_\mathrm{t}$} & \colhead{$\alpha$} &  \colhead{$S_\mathrm{m}$} &  \colhead{$\nu_{m}$}  \\
I.D.  & \colhead{} & \colhead{} & \colhead{[Jy]} &  \colhead{[GHz]}  
}
\startdata
C1 & \nodata      &  $\le -1.6$   & $\le 6.0$      & $\ge27.4$    \\
   & \em 2.5      & $-1.8\pm0.1$  & $3.8\pm0.1$    & $32.0\pm0.1$ \\
\hline
C2 & $0.7\pm0.1$  & \nodata       & $>2.9$         & $>86$        \\ 
   &              & \nodata       & {\em 3.8}      & {\em 115}    \\
\hline
J6 & 1.2\,--\,2.0 & $\ge -1.5$    & 0.90\,--\,0.95 & 65\,--\,119  \\
   & $1.4\pm0.3$  & {\em $-$0.7}  & $0.9\pm0.1$    & $71.4\pm8.0$ \\ 
\hline
J5 & \nodata      & $-0.8\pm 0.1$ & $>$0.5         & $<$15        \\
   & $0.2\pm0.1$ & {\em $-$0.8}  & $0.3\pm 0.1$   & $8.7\pm0.6$  \\
\enddata
\tablecomments{Spectral parameters of selected components. For each component, general constraints are given in the first row and best model parameters are in the second row. For the model parameters, italics denote assumed values.\\ Column designations: $\alpha_\mathrm{t}$, $\alpha$ -- spectral indices of the optically thick and thin parts of the spectrum, respectively; $S_\mathrm{m}$ -- turnover flux density; $\nu_\mathrm{m}$ -- turnover frequency.  For the component C2, the assumed $S_\mathrm{m}$ is set equal to that of the component C1, and the corresponding value of $\nu_\mathrm{m}$ is calculated.}
\end{deluxetable}
\setlength\tabcolsep{6pt}

\subsection{Magnetic fields and equipartition Doppler factors}
\label{sc:bfield}

The estimated turnover frequencies and flux densities of the jet components can be used for calculating their respective magnetic field strengths and Doppler boosting factors. The magnetic field strength of a spherical emitting region with an SSA spectrum is given by \citep{1983ApJ...264..296M}:

\begin{equation}
 B_\mathrm{ssa}[\mathrm{G}] = 10^{-5} b(\alpha) \, \theta_\mathrm{m}^{4} \nu_\mathrm{m}^{5} S_\mathrm{m}^{-2}
 \dfrac{\delta_\mathrm{j}}{1+z} \,,
\label{eq:b_SSA}
\end{equation}
where $b\left(\alpha\right)$ is a coefficient depending on the spectral index (with $b(\alpha)=3.2$ for $\alpha = -0.5$; see Table 1 in \citealt[][]{1983ApJ...264..296M} and Appendix~A in \citealt{2019MNRAS.482.2336P}), $\nu_\mathrm{m}$ is the spectral turnover frequency in GHz, $S_\mathrm{m}$ is the spectral turnover flux density in Jy, $\theta_\mathrm{m}$ is the diameter of the emitting region in mas, and $\delta_\mathrm{j}$ is the Doppler boosting factor. 

The turnover parameters, $S_\mathrm{m}$ and $\nu_\mathrm{m}$, are taken from the spectral fits in Tab.~\ref{table:comp_spix}. The diameters of the emitting regions should be obtained from measurements made at the turnover frequency. In lieu of such measurements, we estimate these diameters by requiring them to reconcile the fitted turnover parameters with the maximum value of brightness temperature listed for a given feature in Tab.~\ref{table:knot_parameters}. 
For this purpose, we first use Eq.~\ref{eq:Tb} which approximates the brightness distribution by a two-dimensional Gaussian and provides its FWHM. To obtain an estimate of $\theta_\mathrm{m}$, the FWHM values are further multiplied by a factor of $\sqrt{3}/\sqrt{2\ln 2}$ accounting for the conversion from the Gaussian to the spherical shape. The conversion factor is determined by calculating the diameter of a sphere filled with homogeneous, optically thin plasma such that it provides the same total and peak flux density as those derived for a given two-dimensional Gaussian component of the modelfit. The resulting estimates of $\theta_\mathrm{m}$ and $B_\mathrm{ssa}$ are listed in the first two columns of Tab.~\ref{table:b-fields}.

Independent estimates of magnetic field strength can be obtained from the equipartition condition \citep{1970ranp.book.....P,2014MNRAS.445.1321Z}, which yields, with the same units as in Eq.~\ref{eq:b_SSA}:
\begin{equation}
B_\mathrm{eq} = 10^{-4} \left[\frac{(1+k_\mathrm{u})\, c_{12}\, \kappa_{\nu}}{f} \frac{S_\mathrm{m}\, \nu_\mathrm{m}}{\theta^{3}_\mathrm{m}\,D_\mathrm{L,Gpc}} 
\frac{(1+z)^{10}}{\delta_\mathrm{j}^4} \right]^{2/7}\,,
\label{eq:b_eq}
\end{equation}
where $D_\mathrm{L,Gpc}$ is the luminosity distance in Gpc, $f$ is volume filling factor of the emitting plasma (with $f$=$1$ adopted hereafter), $k_\mathrm{u}$ is the ratio of total energy in the emitting region to that carried by the SSA population of electrons (with $k_\mathrm{u}\sim 1$ representing an electron-positron flow), the coefficient $c_{12}$ is given in \cite{1970ranp.book.....P}, and $\kappa_\nu$ depends on the low and high frequency cutoffs, $\nu_\mathrm{min}$ and $\nu_{\max}$, of the SSA emission.
For $\nu_\mathrm{min}=10^7$\,Hz and $\nu_\mathrm{max}=10^{13}$\,Hz assumed in this paper, the $\nu_\mathrm{min}$ contribution to $\kappa_{\nu}$ can be neglected, and $\kappa_{\nu}  \approx (\nu_\mathrm{max}/\nu_\mathrm{m})^{1+\alpha}/(1+\alpha)$. The resulting estimates of $B_\mathrm{eq}$ are shown in Tab.~\ref{table:b-fields}.

Both, $B_\mathrm{ssa}$ and $B_\mathrm{eq}$ depend on the Doppler boosting factor, and this dependence can be used for deriving the equipartition Doppler boosting factor, $\delta_\mathrm{eq}$, by requiring that $B_\mathrm{ssa} = B_\mathrm{eq}$. This condition yields
$\delta_\mathrm{eq} \propto (B_\mathrm{SSA}/B_\mathrm{eq})^{7/15}$. Using this approach, we calculate $\delta_\mathrm{eq}$ and present them in Tab.~\ref{table:b-fields} for electron-positron ($k_\mathrm{u}= 1$) and electron-proton \citep[$k_\mathrm{u} = 100$; see][]{2017APh....90...75M} jet.

The observed proper motions of the jet components in \oj at 15\,GHz \citep{2009AJ....138.1874L,2016AJ....152...12L} correspond to apparent speeds $\beta_\mathrm{app}\le 15\,c$, with a median speed of 4.5\,$c$, while at 43\,GHz VLBA-BU-BLAZAR monitoring have estimated values between $\sim$12c and $\sim$7c (Weaver et al. 2021, submitted). One should therefore expect to find Doppler factors $\delta_\mathrm{j} \ge \sqrt{1+\beta^{2}_\mathrm{app}}\!\approx\!5-15$. With this estimate, it is feasible to conclude that plasma condition in the features J6 and J5 are likely to be close to the equipartition. The components C1 and C2 in the core region deviate from the equipartition, with a good indication for this deviation to progressively increase at smaller separations from the origin of the jet.

\begin{deluxetable}{l|rrrrr}
\centering
\tablecaption{Estimates of magnetic field and Doppler factors in the jet.}
\label{table:b-fields}
\tablewidth{0pt}
\tablehead{
 & \mc{$\theta_\mathrm{m}$} & \mc{$B_\mathrm{ssa}$} & \mc{$B_\mathrm{eq}$} & \mc{$\delta_\mathrm{eq}$} & \mc{$\delta_\mathrm{eq}$}\\
    & [mas]&  \mc{[G]} & \mc{[G]}      & \mc{($e^{-}e^{+}$)} & \mc{($e^{-}p$)}
    }
    \startdata
 C1 & 0.04 &   0.002\,$\delta_\mathrm{j}$  &    53\,$\delta_\mathrm{j}^{-8/7}$  & 103   & 173    \\
 C2 & 0.02 & $>$0.011\,$\delta_\mathrm{j}$ & $>$81\,$\delta_\mathrm{j}^{-8/7}$  & $<$63 & $<$105 \\
    &      &   0.027\,$\delta_\mathrm{j}$  &    92\,$\delta_\mathrm{j}^{-8/7}$  & 44    & 75 \\
 J6 & 0.03 &    0.24\,$\delta_\mathrm{j}$  &    39\,$\delta_\mathrm{j}^{-8/7}$  &  11   &  18    \\
 J5 & 0.20 &    0.13\,$\delta_\mathrm{j}$  &   4.1\,$\delta_\mathrm{j}^{-8/7}$  & 5.0   &  8.1    \\
\enddata
\tablecomments{Column designations: $\theta_\mathrm{m}$ -- estimated effective diameter of the emitting region; $B_\mathrm{ssa}$ -- magnetic field strength for synchrotron self-absorbed spectrum; $B_\mathrm{eq}$ -- strength of the equipartition magnetic field; $\delta_\mathrm{eq}$ -- equipartition Doppler factors calculated for electron-positron ($e^{-}e^{+}$, $k_\mathrm{u}=1$) and electron-proton ($e^{-}p$, $k_\mathrm{u}=100$) flow, with $\delta_\mathrm{eq} \propto k_\mathrm{u}^{2/15}$. For the component C2, the second row lists respective values obtained for the assumed spectral turnover point (see Sec. \ref{sc:bfield}).}
\end{deluxetable}

\subsection{Brightness temperature}
\label{sec:tb}

The highest values of $T_\mathrm{b}$ estimated for C1 reach $\approx 5 \times 10^{12}$\,K (see Tab.~\ref{table:knot_parameters}). This is moderately larger than the inverse-Compton limits of $\left(0.3--1.0 \right)\, \times 10^{12}$\,K \citep{1969ApJ...155L..71K} and substantially above the equipartition limit of $\approx 5\times 10^{10}$ \citep{1994ApJ...426...51R}. At the inverse-Compton limit, a jet Doppler boosting factors $\delta_\mathrm{j} \gtrsim 12$ would be required to explain these brightness temperature estimates. Thus, the kinematic constraints $\delta_\mathrm{j}\!\approx\! \left(5-15\right)$ discussed Sec.~\ref{sc:bfield} can reconcile the highest estimated brightness temperature values (see Tab.~\ref{table:knot_parameters}) with the inverse-Compton limit.

\begin{figure}
\centering
\includegraphics[width=0.45\textwidth]{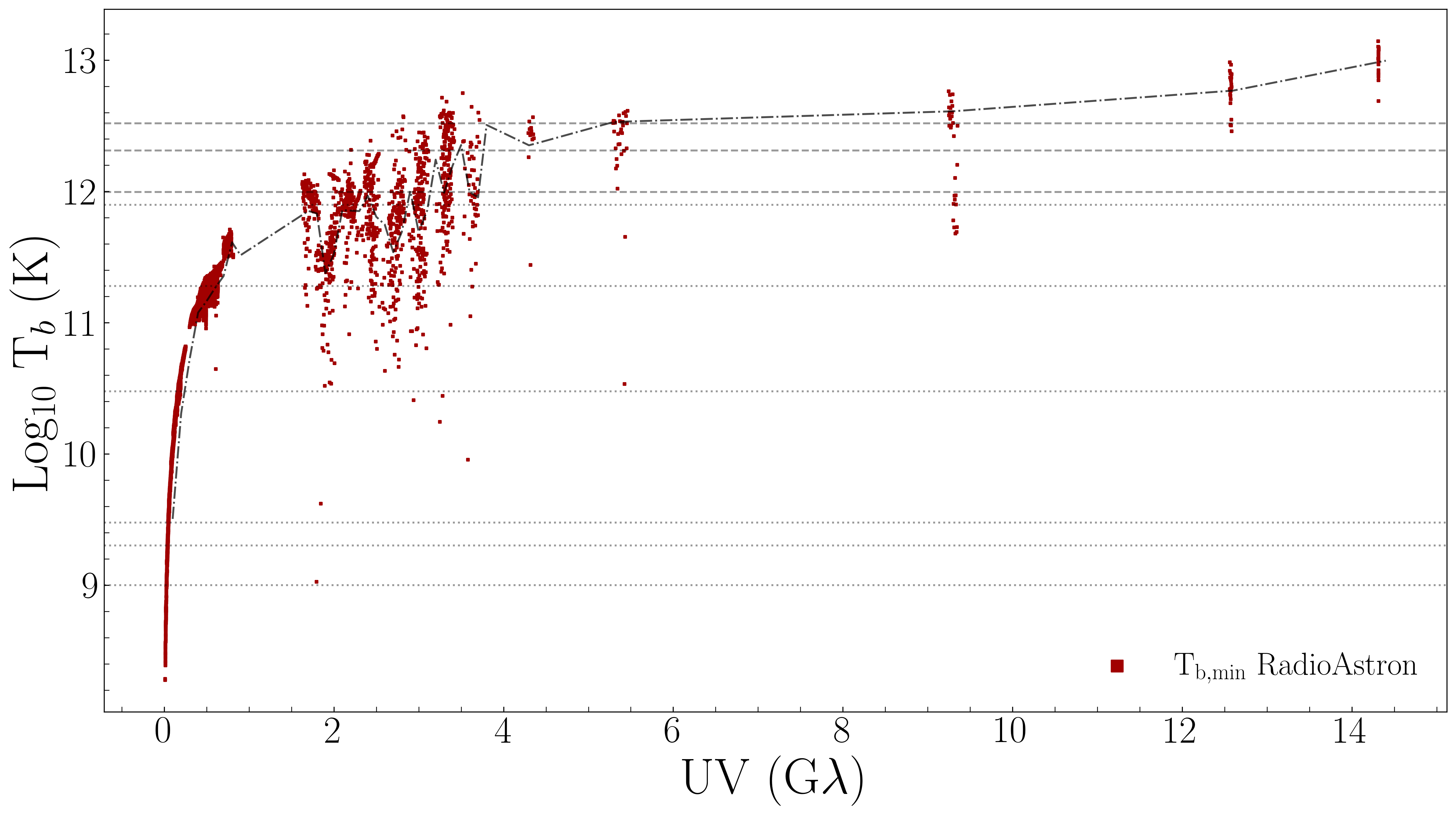}
\caption{Brightness temperatures of modelfit components (horizontal lines) and visibility based estimates (red dots) of minimum brightness temperature, $T_\mathrm{b,min}$ in \oj at 22 GHz.}
\label{fig:ra_tb}
\end{figure}

The 22\,GHz modelfit estimates of $T_\mathrm{b}$ can be compared with the estimates of minimum brightness temperature, $T_\mathrm{b,min}$ made directly from the visibility amplitudes \citep{2015A&A...574A..84L}. This comparison is presented in Fig.~\ref{fig:uvrad_tb} and Fig.~\ref{fig:ra_tb}, showing that $T_\mathrm{b,min}$ reaches $\approx 10^{13}$\,K at the longest $\left(u,u\right)$-spacings coming from the auxiliary long-baseline segments of the observations.

In Fig.~\ref{fig:ra_tb}, $T_\mathrm{b,min}$ shows a continued increasing trend with progressive larger uv-spacing, with no evidence of reaching a plateau. Thus, even larger $T_\mathrm{b,min}$ could in principle be expected at longer baselines than those observed here. \cite{G_mez_2016} report $T_\mathrm{b,min}$ values of the order of $10^{13}$\,K at the longest baselines during \ra observations of the jet in BL\,Lac at 22\,GHz. The authors interpret this extremely high $T_\mathrm{b,min}$ values as resulting from a flaring event that was taking place during the \ra observations, causing the jet flow to depart from equipartition. Similarly, our \oj observations were performed during the onset of a dramatic flaring event in the source (see Sec.~\ref{sec:single-dish}). We should also note that the long baseline snapshots that provide the largest $T_\mathrm{b,min}$ were obtained during different orbits of the SRT, weeks before and after the perigee imaging session, probing therefore probably different flaring states of the source. This provides a natural explanation for why in Fig.~\ref{fig:RA-radplot} the visibilities obtained during the long-baseline snapshots do not provide a good fit to the CLEAN model obtained using only the perigee imaging session.

The measured $T_\mathrm{b,min}\sim 10^{13}$\,K require Doppler boosting factors, $\delta_\mathrm{j}$, of the order of 10\,--\,30 to reconcile them with the inverse-Compton limit. This is broadly in agreement with the values estimated from the proper motion of components moving downstream the jet by the MOJAVE \citep{2016AJ....152...12L} and VLBA-BU-BLAZAR (Weaver et al., 2021, submitted) monitoring programs of $\delta_\mathrm{j}\!\approx\!5-15$ and $\delta_\mathrm{j}=8.6\pm2.8$, respectively. At the upper boundary of this range, the respective viewing angle of the jet should be $\theta_\mathrm{j}\!\approx\!3^\circ-8^\circ$. Reducing this angle by about 2$^\circ$ would lead to increasing the Doppler boosting factor up to $\sim30$ even without requiring the bulk Lorentz factor to increase.

\subsection{Faraday rotation}
\label{subsec:rm}

\begin{figure*}
\includegraphics[width=\textwidth]{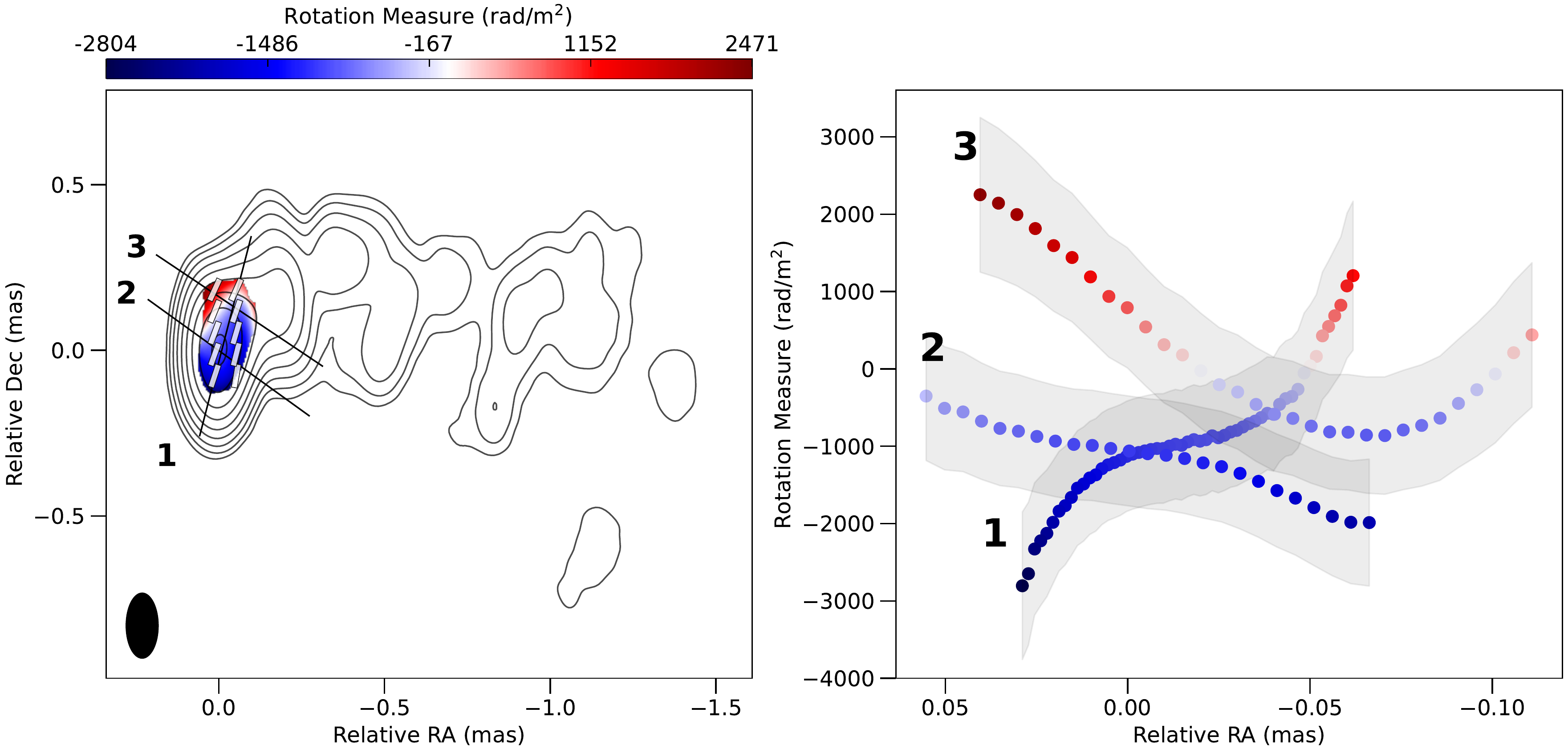}
\caption{\textit{Left:} rotation measure map obtained from the 22, 43, and 86 GHz images (convolved with the same 0.2$\times$0.1 mas FWHM beam). Contours correspond to the 22 GHz image and white bars to the Faraday-corrected EVPAs. \textit{Right:} rotation measure values along the three cuts displayed on the left panel, starting from the labeled numbers. Shaded regions indicate the profiles uncertainty.}
\label{Fig:RM}
\end{figure*}

According to theoretical models, jet launching from super-massive black holes is electromagnetic in nature through the action of helical magnetic fields, either anchored in the accretion disk or the black hole ergosphere \citep{10.1093/mnras/179.3.433,1982MNRAS.199..883B}. The observational signature of these fields is imprinted in the polarization information of the synchrotron radiation, and is tightly connected to the phenomenon of Faraday rotation: the rotation of the polarization plane that occurs when a polarized electromagnetic wave passes through a magnetized plasma (the so-called Faraday screen). The rotation of the polarization angle, $\chi$, introduced by the Faraday screen is determined as $\chi = \chi_{o} + \rm{RM} \times \lambda^{2}$, where $\lambda$ is the observed wavelength and $\chi_{0}$ is the intrinsic electric vector position angle of the emitting region. The rotation measure (RM) is expressed by \citep[e.g.,][]{2017isra.book.....T}:

\begin{equation}
    \rm{RM} = \frac{e^{3}}{8 \pi^{2} \epsilon_{0} m_{e}^{2} c^{3}} \int \nu_{e} \textbf{B}_{\parallel} d\textbf{l},
    \label{chap1_eq:RM2}
\end{equation}

\noindent
where RM is measured in $\rm{rad/ cm}^{2}$, $\epsilon_{0}$ is the permittivity of the vacuum, $\nu_{e}$ is the electron density, $\textbf{B}_{\parallel}$ is the component of the magnetic field which is parallel to our line of sight, and $d\textbf{l}$ is the path length from the source to the observer through the de-polarizing plasma. The sign of the Faraday rotation coincides with the sign of the line-of-sight magnetic field.

In this work, we have combined our highest resolution polarimetric images obtained with the ground array at 43 and 86\,GHz, with that of \ra at 22\,GHz (natural weighted image) to produce the RM map of the innermost jet regions in \oj. For this we first convolved the images at the three different frequencies with a common restoring beam of $0.2\times0.1$\,mas at position angle of $0\degr$, which slightly over-resolves the 7\,mm image while still preserving a fraction of the higher resolution achieved at 86\,GHz and 22\,GHz. Alignment of the images at the three different frequencies was obtained by performing a cross-correlation of the total intensity images following the approach described in \cite{2016ApJ...817...96G}, and references therein. The estimated shifts were all of the order of the image pixel size of 6$\mu$as, and therefore no corrections were applied.

The obtained RM map, with overlaid Faraday-corrected EVPAs is shown in Fig.~\ref{Fig:RM}. It should be noted that the images used to compute the RM were obtained at different epochs (see Tab.~\ref{table:obs}) during a flaring state of the source (see Sec.~\ref{sec:single-dish}), and therefore the reliability of the obtained RM map relies on the assumption of negligible polarization structural changes in the time span covering the considered observations.

Figure \ref{Fig:RM} reveals a region with enhanced RM in the VLBI core area of \oj, with a median RM of $-1000\pm300$ rad/m$^2$. This is broadly consistent (in magnitude) with a previous core RM estimation of $-$367$\pm$71 rad/m$^2$ based on simultaneous VLBI observations between 8\,GHz and 15\,GHz taken in 2006 April 28 \citep{2012AJ....144..105H}.

The size of the RM region is of the order of our resolution beam, and therefore we lack the necessary angular resolution to accurately measure gradients in the RM \citep[e.g.,][]{2010ApJ...722L.183T}. With this limitation in mind, we have plotted in the right panel of Fig.~\ref{Fig:RM} several cuts of the RM perpendicular and parallel to the local jet direction, which are indicative for the presence of gradients across and along the jet. Gradients in RM along the jet direction are expected to be associated with a progressive decrease in the magnetic field strength and electron energy density with distance along the jet \citep{Jorstad_2007}. On the other hand, gradients across the jet width are indicative for the presence of a toroidal magnetic field component. This is also consistent with the measured Faraday-corrected EVPAs, which are aligned with the local jet direction, suggesting also the presence of a predominant toroidal component in the magnetic field. The negative gradient in RM from north-east to south-west direction suggests a toroidal magnetic field oriented clockwise as seen in the direction of flow motion.

The indicative RM gradient across the jet width and EVPAs are both in agreement with the presence of a helical magnetic field threading the innermost regions of the jet in \oj, as predicted by jet formation models \citep{Blandford:1993tz,1982MNRAS.199..883B} and observed previously in a number of sources \citep[e.g.,][]{denise2004,2009MNRAS.393..429O,Hovatta:2012jv,galaxies5010011}.

\section{Discussion and Conclusions}
\label{sec:discussion}

\oj is considered as one of the best candidates for harboring a binary black hole (BBH) system in its center. The updated BBH central engine description for \oj allows us to track the changes in the orientation of the accretion disk and the primary BH spin \citep{2013A&A...557A..28V}. However, additional assumptions are needed to explain its decades long radio jet observations. Interestingly, the assumption that the jet is launched perpendicular to the innermost disk axis leads to BBH orbital time scale observational implications \citep{2012MNRAS.427...77V,2021MNRAS.503.4400D}. Further, the time evolution of the innermost disk axis shows up as a bending of the radio jet. This is because the changes at the launch angle are propagated to the more distant parts of the jet with a time delay. The rather small variations of the launch angle are further amplified by projection effects since the viewing angle of the jet is small. This is in agreement with our observations that confirm a progressive jet bending with increasing angular resolution up to the smallest spatial scales probed by \ra.

Employing VLBI data sets at 15, 43 and 86\,GHz, a consistent description for the temporal evolution of \oj radio jet was provided in \cite{2021MNRAS.503.4400D} making use of a helicity parameter that allows for outward jet motion that is not exactly in a straight line, as noted in \cite{2013A&A...557A..28V}. In addition, one may use the information of the time evolution of optical polarization and determine the jet orientation close to the jet launch site \citep{2012MNRAS.421.1861V,sasada18}.

In the jet distance range from 0.2\,mas to 1\,mas the component position angles (PA) listed in Tab.~\ref{table:knot_parameters} are in agreement with \cite{2021MNRAS.503.4400D} within the errors. For example, the PA of component J6 is $-55\pm10^{\circ}$ in the model while here we find $-36\pm13^{\circ}$ (86\,GHz). The optical polarization data arise from a knot at unknown distance, but judging from the agreement of the PA of our innermost knot C2 of the 86\,GHz map and the PA of the model \citep{2012MNRAS.421.1861V}, it is possible that the optical emission region is not far from knot C2.
Beyond 1\,mas our PA values agree with the earlier model of \cite{2012MNRAS.421.1861V}.

From our BBH central engine prescription, we expect the inner components C2 and J6 to rotate counter clockwise (i.e., increasing values of PA) by about $15^{\circ}$ degrees between 2014 and 2017 while the changes further out in the jet would be barely noticeable.

Even though the observed periodicity in the light curve of \oj and innermost position angle changes can be explained by a BBH model, alternative physical models can also lead to a similar phenomenology. 

\cite{2018MNRAS.478.3199B}, based on 22 years observations at 15\,GHz found that the jet of \oj is rotating with a period of $\sim$30 yr. Modeling of \oj radio data showed that this rotation can be explained by a combination of jet precession and nutation. The physical cause of the precession can be driven by a binary system in \oj center, as well as by the mechanism of the Lense-Thirring precession by the tilted accretion disk of a single BH \citep[e.g.,][]{2018MNRAS.474L..81L}, which in the case of \oj provides realistic parameters.

\cite{2012ApJ...747...63A} propose that variable asymmetric injection of the jet flow, perhaps related to turbulence in the accretion disc, coupled with hydrodynamic instabilities leads to the non-ballistic dynamics that causes the observed non-periodic changes in the direction of the inner jet. \cite{galaxies5010012} presented evidence that \oj is behaving as a rotating helix based on the study of MOJAVE 15\,GHz VLBA images from 1995 to 2015. The results of the ridge line analysis of the data showed that the jet is rotating with a period of possibly $\sim$30 yr. The inner jet apparently seems to have moved to a new direction after the rotation, indicating that jet nozzle has been re-oriented. A model of a helical jet, observed from a small and varying viewing angle, had been proposed earlier by \cite{2013A&A...557A..28V}. Another suggested scenario by \cite{2017A&A...597A..80H} proposes that changes in the position angle of the jet are due to opacity shifts of the observed core in a bent jet.

On the other hand, the 12-year periodicity of \oj optical light-curves, that usually is attributed to a binary black hole system in the center of \oj can be also explained by the existence of a non-radially moving feature along a helical jet. \cite{2020Univ....6..191B}, showed that the differences between the peak flux values of the periodic optical flares, as well as the time-lag between optical and radio repeated variability, can be caused by the development of helical mode of the Kelvin-Helmholtz instability, inside a, well-aligned with our line of sight, helical jet.

Our polarization images are consistent with the innermost jet in \oj being threaded by a predominantly toroidal magnetic field, while our Faraday rotation map is indicative of the presence of a gradient in rotation measure across and along the jet. Both pieces of information suggest that the innermost jet region in \oj is threaded by a helical magnetic field, as expected from jet formation models, and in agreement with previous studies. In particular \cite{2018A&A...619A..88M} report a clockwise EVPA rotation by $\sim\!340^{\circ}$ during a multi-frequency single-dish campaign taken with the 100\,m Effelsberg radio telescope in 2016, which is interpreted as produced by a polarized component propagating on bent jet threaded by a helical magnetic field.

The spectral analysis combining our \ra observations and ground-based VLBI observations is in agreement with the parsec-scale jet being in equipartition between the particles and magnetic field. However, there are some clear evidence for the jet being dominated by the internal energy of the emitting particles as we probe progressively closer to the central engine in the VLBI core area, in agreement with the onset of a large multi-wavelength flare that peaked few months after our VLBI observations.

Brightness temperatures have been estimated from the model fitted components, as well as from the visibility amplitudes. The maximum observed brightness temperature of $5.2\times10^{12}$\,K for the VLBI core can be reconciled with the inverse-Compton limit assuming a moderate Doppler boosting factor of the order of 5--15, in agreement with those estimated from the proper motion of superluminal components. Ground--space fringes have been detected up to a record projected baseline distance of 15.1 Earth diameters in length (one of the longest ever obtained with \ra at 22\,GHz), from which we have estimated a minimum brightness temperature of $T_\mathrm{b,min}\!\sim\!10^{13}$\,K. The rising $T_\mathrm{b,min}$ trend with projected baseline length with no indication of reaching a plateau suggests that even larger brightness temperatures could be measured with longest baselines, although such extremely high brightness temperatures could be explained by larger Doppler boosting factors than those measured at parsec-scales if the innermost jet is pointing close to our line of sight. Alternatively, they may be an indication for the presence of other physical phenomena, as demonstrated by the \ra observations of the quasar 3C\,273 from \cite{2016ApJ...820L...9K}.

Further \ra observations of \oj have been performed in 2016, 2017, 2018, which are also in combination with quasi-simultaneous ground-based millimeter VLBI observations taken with the Event Horizon Telescope (in 2017 and 2018) at 230\,GHz, as well as with the Global Millimeter VLBI Array (GMVA), including phased-ALMA, at 86\,GHz. These observations, together with accompanying multi-wavelength  coverage \citep{2020MNRAS.498L..35K,2021MNRAS.504.5575K} have the potential to either spatially resolve the hypothetical binary black hole system in \oj, or to put strong constrains on binary black hole and alternative models that predict the changing innermost jet position angle and overall periodic flaring activity that characterize this enigmatic source.


\section{Acknowledgements}

The Work at the IAA-CSIC is supported in part by the Spanish Ministerio de Econom\'{\i}a y Competitividad (grants AYA2016-80889-P, PID2019-108995GB-C21), the Consejer\'{\i}a de Econom\'{\i}a, Conocimiento, Empresas y Universidad of the Junta de Andaluc\'{\i}a (grant P18-FR-1769), the Consejo Superior de Investigaciones Cient\'{\i}ficas (grant 2019AEP112), and the State Agency for Research of the Spanish MCIU through the Center of Excellence Severo Ochoa award for the Instituto de Astrof\'{\i}sica de Andaluc\'{\i}a (SEV-2017-0709).
YYK, PAV, ABP acknowledge support from the Russian Science Foundation grant 21-12-00241.
The RadioAstron project is led by the Astro Space Center of the Lebedev Physical Institute of the Russian Academy of Sciences and the Lavochkin Scientific and Production Association under a contract with the Russian Federal Space Agency, in collaboration with partner organizations in Russia and other countries. 
The European VLBI Network is a joint facility of of independent European, African, Asian, and North American radio astronomy institutes. Scientific results from data presented in this publication are derived from the EVN project code GA030E.
This research is partly based on observations with the 100~m telescope of the MPIfR at Effelsberg. 
The VLBA is an instrument of the National Radio Astronomy Observatory, a facility of the National Science Foundation operated under cooperative agreement by Associated Universities. 
This research has made use of data obtained with the Global Millimeter VLBI Array (GMVA), which consists of telescopes operated by the MPIfR, IRAM, Onsala, Metsahovi, Yebes, the Korean VLBI Network, the Green Bank Observatory and the National Radio Astronomy Observatory.
This publication makes use of data obtained at Mets\"ahovi Radio Observatory, operated by Aalto University in Finland.
Our special thanks to the people supporting the observations at the telescopes during the data collection.
This research is based on observations correlated at the Bonn Correlator, jointly operated by the Max-Planck-Institut f\"ur Radioastronomie (MPIfR), and the Federal Agency for Cartography and Geodesy (BKG). 
This research has made use of data from the MOJAVE database that is maintained by the MOJAVE team (Lister et al. 2018). 
This study makes use of 43~GHz VLBA data from the VLBA-BU Blazar Monitoring Program (VLBA-BU-BLAZAR; \url{http://www.bu.edu/blazars/VLBAproject.html}), funded by NASA through the Fermi Guest Investigator grant 80NSSC20K1567. \\

\software{AIPS, \cite{1990apaa.conf..125G}, DiFX, \cite{2016Galax...4...55B}, PIMA, \cite{2011AJ....142...35P},  DIFMAP, \cite{Shepherd:1997wv} }

\facility{\emph{RadioAstron} Space Radio Telescope (\textit{Spektr-R}), EVN, GMVA, VLBA, OVRO, RATAN-600, Effelsberg, Mets\"{a}hovi Radio Telescope, SMA, IRAM-30m.}

\bibliography{Referencias,OJ287}{}

\clearpage
\appendix

\section{Calibration of the instrumental polarization}
\label{app:B}

\subsection{RadioAstron data}
\label{Sec:RA_pol_cal}

\begin{wrapfigure}{r}{0.5\textwidth}
\includegraphics[width=0.48\textwidth]{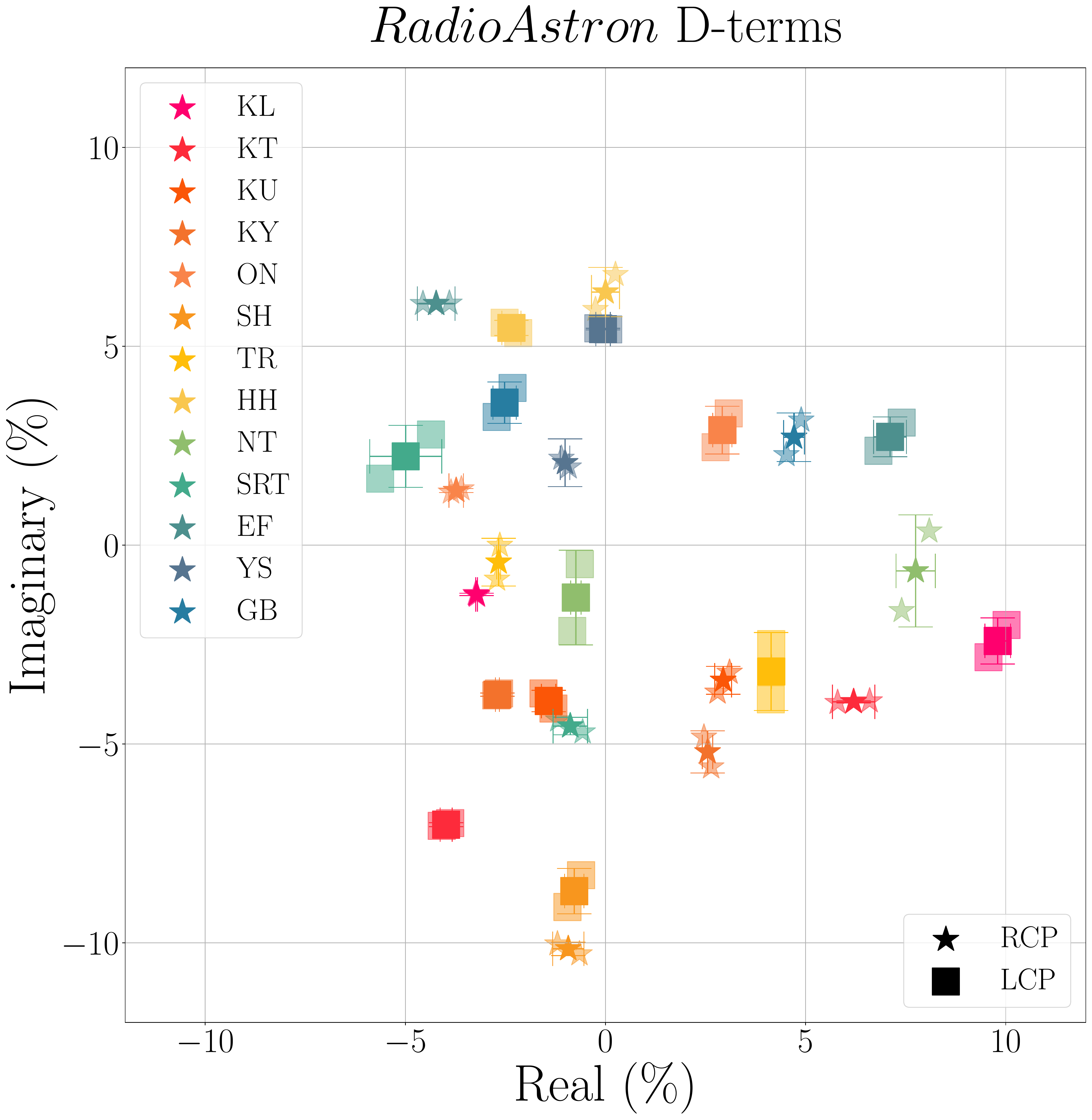}
\caption{D-Terms for the RadioAstron perigee observations. Plotted values correspond to each IF and the median, with errors estimated from the dispersion in values across both IFs.}
\label{Fig:RA-Dterms}
\end{wrapfigure}

AIPS's task LPCAL \citep{Leppanen:1995eg} was used to estimate the instrumental polarization leakage, also known as D-terms, independently for each IF. The target source \oj was used for computing the D-terms since it was the only source observed simultaneously with the SRT while providing the best parallactic angle coverage for the ground antennas. Fig.~\ref{Fig:RA-Dterms} shows the obtained D-terms for all the participating antennas during the perigee imaging session. We find consistent values across both IFs, confirming the reliability of the estimated D-terms. For the SRT we obtained $(-0.88-4.55j)\pm(0.30+0.15j)$ and $(-4.99+2.23)\pm(0.63+0.55j$), for RCP and LCP respectively. The dispersion across the two IFs has been used to estimate the errors.

Absolute calibration of the EVPA was obtained through comparison with VLA observations of \oj at 22.295~GHz performed in 2014 May 1 under a different observing program (Marscher et al., private communication). Given the time span between our \ra observations and the VLA ones we estimate an error for the absolute EVPA calibration, $\Delta \chi$, of the order of 10$^{\circ}$.

\subsection{GMVA polarization calibration}
\label{Sec:Polarization_calib}

Similarly to the \ra observations, the calibration of the instrumental polarization leakage for the 86\,GHz data was performed by employing AIPS' task LPCAL.

LPCAL assumes that the source structure can be described as a collection of polarization components each one with a constant fractional polarization, known as the self-similarity assumption. A good parallactic angle coverage is also required for the LPCAL fitting algorithm. Hence, ideally LPCAL should be used on sources with a simple polarization structure (preferably with a low fractional polarization), and a good parallactic angel coverage, requirements that are rarely met in millimeter VLBI observations. Alternative methods should then be considered when these requirements are not fully met. For this reason we have tested three different approaches to compute the D-terms for the 86\,GHz GMVA observations.

\begin{figure*}
\centering
\includegraphics[width=0.4\textwidth]{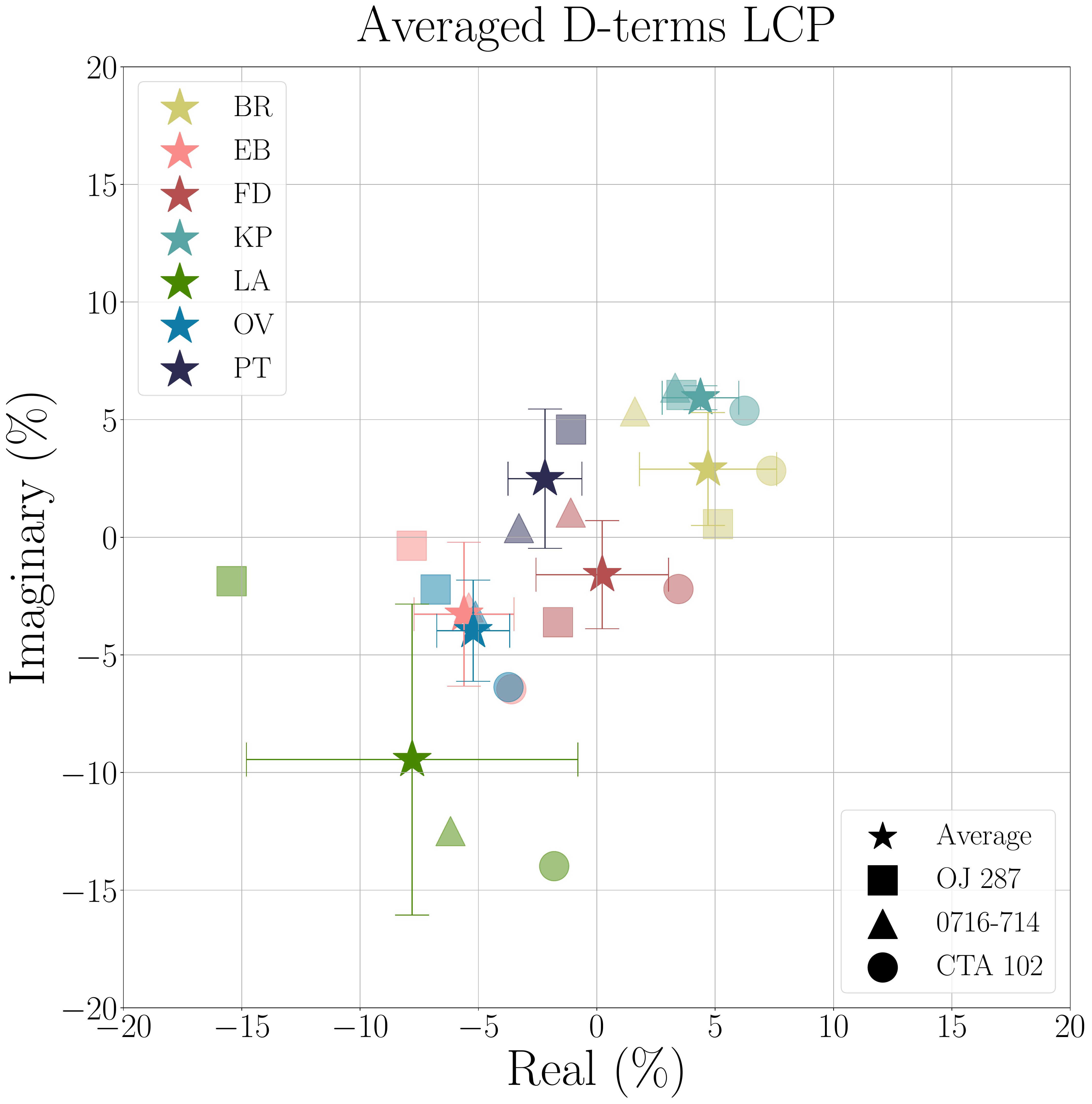} 
\includegraphics[width=0.4\textwidth]{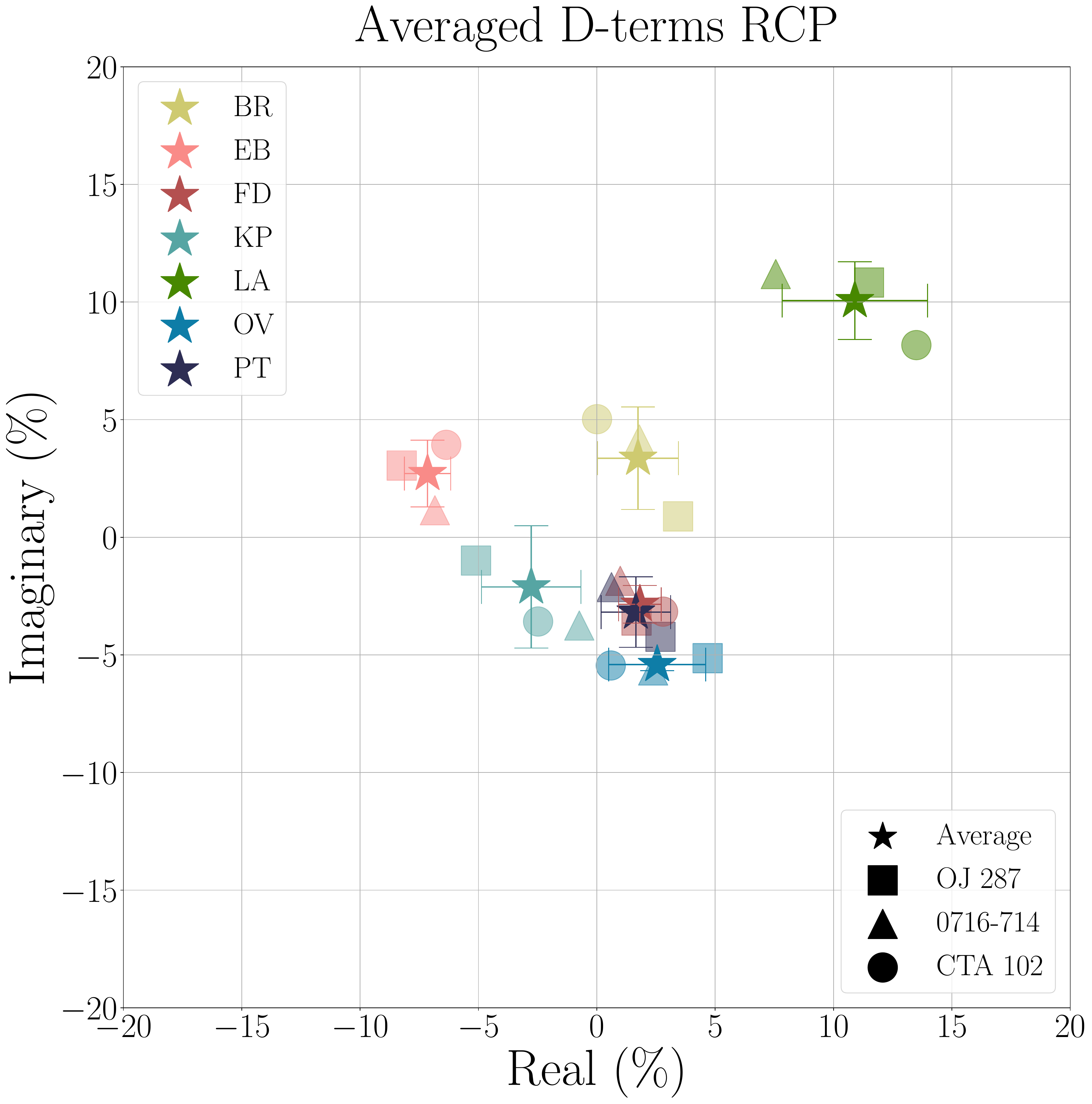} 
\caption{D-terms derived based on the data for \oj, 0716+714, and CTA\,102, and their mean value, for the two polarizations. Labels mark the different radio telescopes. The uncertainties are estimated based on the standard deviation of the individual D-terms values.}
\label{fig:gmva_dterms}
\end{figure*}

First we computed the D-terms using \oj, which provided consistent values across all IFs, with a small dispersion of the order of $\sim$2\% in amplitude and 13$^{\circ}$ in phase. Following \cite{2019A&A...622A.158C}, we also computed the D-terms for each one of the bright and compact sources that were observed in the same session together with \oj and had a large parallactic angle coverage ($\geq$80$^{\circ}$), taking the median values between the different intermediate frequency channels for each antenna as the representative D-terms. Fig.~\ref{fig:gmva_dterms} shows the obtained D-terms for each source, including our target, and the median values. Finally we also tested which one of the D-terms obtained for each individual source provided the highest polarization image dynamic range across all observed sources. Out of these tests we found that the D-terms provided by \oj yielded the polarization image with the highest dynamic range.

Calibration of the absolute EVPA for the GMVA 86\,GHz observations was carried out through comparison with single dish IRAM 30 m telescope of \oj as part of the POLAMI monitoring program, with an estimated uncertainty of the order of $5^{\circ}$ \citep{2018MNRAS.474.1427A,2018MNRAS.473.1850A}.

\subsection{Estimations and uncertainties of the polarimetric parameters}
\label{subsec:polarization}

The polarized flux density in this work is computed for all data sets by the standard formula $P=\sqrt{Q^{2}+U^{2}}$, whereas, the uncertainty of $P$, $\sigma_\mathrm{P}$, is estimated by taking into account a calibration uncertainty of about 10\% of the polarized flux density and a statistical error provided by the map thermal noise \citep{2014A&A...571A..54L} as $\sigma_{P}=\sqrt{ \left( 0.1 \times P \right)^{2} + rms_\mathrm{P}^{2} }$, based on the polarization flux estimations that were obtained from the image domain of each image.

The fractional polarization percentage is determined as $m= 100 \left( P/S \right)$, and the error on $m$ is given by: 

\begin{equation}
\Delta_{m}= \frac{1}{S} \sqrt{\sigma_{P}^{2} + \left( m \times \sigma_{S} \right)^{2} + \sigma_\mathrm{D-term}^{2}}
\end{equation}

\noindent
The term $\sigma_\mathrm{D-term}$ represents the systematic polarization calibration error, which is defined as \citep{1994ApJ...427..718R, 2012AJ....144..105H}: 

\begin{equation}
\sigma_\mathrm{D-term}=\sigma_\mathrm{amp} \frac{\sqrt{S_\mathrm{total}^{2}+ \left( 0.3 \times S_\mathrm{peak}\right)^{2} }}{\sqrt{N_\mathrm{ant} N_\mathrm{IF} N_\mathrm{scan}} }
\end{equation}

\noindent
where $\sigma_{amp}$ is the standard deviation of the D-term amplitudes, $N_\mathrm{ant}$ is the number of antennas, $N_\mathrm{IF}$ is the number of the IFs, and $N_\mathrm{scan}$ is the number of independent scans with different parallactic angles. 

For the GMVA image we have $\sigma_{amp}\sim 2\%$ , $N_\mathrm{ant}=7$, $N_\mathrm{scan}=5$, $N_\mathrm{IF}=16$, $S_\mathrm{peak}=3.47$\,Jy, and $S_\mathrm{total}=5.07$\,Jy, which results in $\sigma_\mathrm{D-term}=4.62$\,mJy/beam. Similarly, for the \textit{RadioAstron} data we have $\sigma_{amp}\sim 1\%$, $N_\mathrm{ant}=13$, $N_\mathrm{scan}=6$, $N_\mathrm{IF}=2$, $S_\mathrm{peak}=1.18$\,Jy, and $S_\mathrm{total}=2.38$\,Jy, which results in $\sigma_\mathrm{D-term}=1.92$\,mJy/beam. For the 15\,GHz data, $\sigma_\mathrm{D-term}\approx 2\times rms_{P}$ and the error for the absolute EVPA calibration is of the order of $5^{\circ}$ \citep{2012AJ....144..105H}, whereas for the 43\,GHz data, $\Delta m=1$\% and $\Delta \chi=5^{\circ}$\% \citep{2005AJ....130.1418J}.

\end{document}